\documentclass[12pt]{article}

\usepackage{mathrsfs}
\usepackage{array}
\usepackage{amssymb}
\usepackage{amsmath}
\usepackage{graphicx}
\newcommand{\p}{\partial}
\newcommand{\LL}{\mathscr{L}}

\newcommand{\e}{\epsilon}

\newcommand{\be}{\beta}
\newcommand{\na}{\nabla}

\newcommand{\mR}{\mathbb{R}}

\def\e{\epsilon}

\def\be{\begin{equation}}
\def\ee{\end{equation}}
\def\bea{\begin{eqnarray}}
\def\eea{\end{eqnarray}}
\def\makeatletter{\catcode`\@=11}
\makeatletter
\def\mathbox#1{\hbox{$\m@th#1$}}%
\def\math@ccstyles#1#2#3#4#5#6#7{{\leavevmode
      \setbox0\mathbox{#6#7}%
      \setbox2\mathbox{#4#5}%
      \dimen@ #3%
      \baselineskip\z@\lineskiplimit#1\lineskip\z@
      \vbox{\ialign{##\crcr
             \hfil \kern #2\box2 \hfil\crcr
             \noalign{\kern\dimen@}%
             \hfil\box0\hfil\crcr}}}}
\def\mathaccstyles{\math@ccstyles\maxdimen}
\def\maththroughstyles{\math@ccstyles{-\maxdimen}}
\def\unitmatrixDT%
 {\maththroughstyles{.45\ht0}\z@\displaystyle {\mathchar"006C}\displaystyle 1}

\begin{document}

\begin{titlepage}
\begin{center}

\rightline{UG-12-01}
\rightline{ICCUB-12-307,UB-ECM-PF 12/75}
\rightline{\today}

\vskip 1.5cm

{\Large \bf \hskip -.4truecm  `Stringy'  Newton-Cartan Gravity}

\vskip 1cm

{\bf \hskip -1.1truecm Roel Andringa$\,{}^*$,
                       Eric Bergshoeff\,${}^*$,
                       Joaquim Gomis\,${}^\dagger$ and
                       Mees de Roo\,${}^*$}

\vskip 25pt

{\em \hskip -.1truecm ${}^*$\ Centre for Theoretical Physics, University of
Groningen, Nijenborgh 4, 9747 AG Groningen, The Netherlands\vskip 5pt}

{\em \hskip -.1truecm ${}^\dagger$\ Departament ECM and ICCUB, Universitat
de Barcelona, Diagonal 647, 08028, Barcelona, Spain\vskip 5pt}

{email: {\tt R.Andringa@rug.nl, E.A.Bergshoeff@rug.nl, gomis@ecm.ub.es, M.de.Roo@rug.nl}}

\vskip 15pt

\end{center}

\vskip 1.2cm

\begin{center} {\bf ABSTRACT}\\[3ex]
\end{center}

We construct a ``stringy'' version of Newton-Cartan gravity in which the concept of a Galilean observer plays
a central role. We present both the geodesic equations of motion for a fundamental string and  the bulk equations of motion
in terms of a gravitational potential which is a symmetric tensor with respect to the longitudinal directions
of the string.
The extension to include a non-zero cosmological constant is given. We stress the symmetries and (partial)
gaugings underlying our construction. Our results provide a convenient starting point to investigate
applications of the  AdS/CFT correspondence based on the  non-relativistic ``stringy'' Galilei algebra.



\end{titlepage}

\newpage
\setcounter{page}{1} \tableofcontents

\newpage

\setcounter{page}{1} \numberwithin{equation}{section}

\section{Introduction}

Einstein's special relativity is based upon an equivalence between frames that are connected to each other
by the Poincar\'e symmetries, consisting of translations and Lorentz transformations in a $D$-dimensional spacetime.\,\footnote{Since our arguments do not depend on the dimension we keep the dimension $D$ of spacetime arbitrary.}
The extension to general relativity can be viewed as the gauge theory of these Poincar\'e transformations
where the constant parameters of the different transformations have been promoted to arbitrary funtions
of the spacetime coordinates $x^\mu\ (\mu=0,1,\ldots,D-1)$. This leads to a theory invariant under general coordinate transformations.
In general relativity the curvature of spacetime is described by an invertable metric function
$g_{\mu\nu}(x)$ which is symmetric in the spacetime indices and which replaces the Minkowski metric $\eta_{\mu\nu}$
of flat spacetime corresponding to special relativity. The equations of motion for the metric function are given by the well-known
Einstein's equations of motion which are basically a set of second-order differential equations for $g_{\mu\nu}(x)$ with the energy-momentum tensor
as a source term. The equation of motion of a particle moving in a curved spacetime is given by the geodesic equation
corresponding to  that spacetime. All equations transform covariantly with respect to general coordinate transformations.

One of the observations underlying general relativity is that an observer in a local ``free-falling'' frame does not experience any gravitational force.
Consequently, the equation of motion of a particle in such a frame describes a straight line corresponding to motion with a
constant velocity. These equations of motion transform covariantly under the Poincar\'e symmetries of special relativity. Indeed, locally,
general relativity coincides with special relativity corresponding to $g_{\mu\nu}(x)=\eta_{\mu\nu}$.

To apply general relativity in practical situations it is often convenient to consider the
Newtonian limit which  is defined as the limit of small velocities $v << c$ with respect to the speed of light $c$, and  a slowly varying and weak
gravitational field. The Newtonian limit is not the unique non-relativistic limit of general relativity. It is a specific limit which is based upon the assumption that particles are the basic entities and it further makes the additional assumption of a slowly varying and weak
gravitational field. In this work we will encounter different limits which are based upon strings or, more general, branes, as the basic objects, and
which do not necessarily assume a slowly varying and weak gravitational field.

Taking the Newtonian limit
there is a universal time $t$ and there is only equivalence between frames that are
connected to each other  by the Galilei symmetries, consisting of (space and time) translations, boost transformations
and $(D-1)$-dimensional spatial rotations.
Like in general relativity, an observer in a free-falling frame does not experience any gravitational force. All free-falling frames
are connected to each other by the Galilei symmetries. For practical purposes, it is convenient to consider not only
free-falling frames but to include all frames corresponding to a  so-called ``Galilean observer'' \cite{Kuchar:1980tw,PietriLusannaPauri}. These are all frames
that are accelerated, with arbitrary (time-dependent) acceleration, with respect to a free-falling frame. An example of a
frame describing a Galilean observer with constant acceleration \cite{Lukierski:2007nh} is the one attached to the Earth's surface, thereby ignoring the rotation of the Earth. Newton showed
that in the constant-acceleration frames the gravitational force is described by a time-independent scalar potential $\Phi(x^i)\ (i=1,\cdots, D-1)$, the so-called Newton potential. In frames with time-dependent acceleration the potential becomes an arbitrary function $\Phi(x)$ of the
spacetime coordinates. A noteworthy difference between general relativity and Newtonian gravity is that, while in general relativity \emph{any} observer can locally in spacetime use a general coordinate transformation to make the metric flat, in Newtonian gravity only the Galilean observers can use an acceleration to make the Newton potential disappear.
The Newton potential deforms the free motion of a particle and is itself described by a Poisson equation with
the mass density $\rho(x)$ as a source term, and it takes over the role played by the metric function in general relativity. In the
Newtonian limit the Newton potential is contained in the time-time component of $g_{\mu\nu}(x)$, and the potential term in the geodesic equation is given by the space-time-time component of the Christoffel symbol.

The equations of motion corresponding to a Galilean observer are invariant under the so-called ``acceleration-extended'' Galilei symmetries.
This corresponds to an extension of the Galilei symmetries in which the  (constant) space translations and boost
transformations have been gauged resulting into a theory which is invariant under arbitrary time-dependent spatial translations.\,\footnote{
 The group of acceleration-extended Galilei symmetries is also  called the Milne group \cite{Duval:1993pe}.}
The gravitational potential can be viewed as the ``background gauge field'' necessary to realize these time-dependent translations.
Starting from a free particle in a Newtonian spacetime, there are now two ways to derive the equations of motion for a Galilean observer from a gauging principle. If one is only interested in the physics observed by a Galilean observer it is sufficient to gauge the
constant space translations by promoting the corresponding (constant) parameters to  arbitrary functions of time. This automatically
includes the gauging of the boost transformations. The equation of motion of a particle is then
obtained by deforming the free equation of motion with the background gravitational potential $\Phi(x)$ such that the resulting equation
is invariant under the acceleration-extended Galilei symmetries. The Poisson equation of $\Phi(x)$ can be obtained
by realizing that it is the only equation, of second order in the spatial derivatives, that is invariant under the acceleration-extended
Galilei symmetries.

In case one is interested in not only the physics as experienced by a Galilean observer but by other observers as well, corresponding to, e.g.,
rotating frames, it is convenient to first gauge {\it all}
symmetries of the Newtonian theory. One thus ends up with a gravitational theory invariant under much more symmetries than the acceleration-extended Galilean symmetries. This procedure
has been described in  \cite{PietriLusannaPauri}, and somewhat differently in \cite{Andringa:2010it}. The gauging  contains an additional subtlety with respect to the relativistic case.
In the relativistic case both the
equations of motion and the Lagrangian leading to the equations of motion are invariant under the Poincar\'e symmetries.
This is different from the Newtonian case. It turns out that, although the equations of motion are invariant under the Galilei symmetries, the
corresponding Lagrangian is only invariant under boosts up to a total time derivative. This leads to a central extension of the Galilei algebra,
containing  an extra so-called central charge generator $Z$,
which is called the Bargmann algebra \cite{Bargmann}.\,\footnote{Alternatively, one may construct an invariant
Lagrangian at the expense of introducing an additional coordinate. One thus ends up with a higher-dimensional realization of the Bargmann algebra in which the central charge transformation
corresponds to a translation in the extra direction \cite{Duval:1984cj}.}
The gauging procedure, in order to be well-defined, must be applied to the Bargmann algebra.
Once one decides to restrict to a Galilean observer, with flat spatial directions, one must
impose as kinematical constraints that the curvature with respect to the spatial rotations vanishes.
It should be stressed that one is not forced to impose this curvature constraint, and one could stay more general and try to solve the resulting theory of gravity for a curved transverse space. But if one does restrict to a flat transverse space and a Galilean observer, the gauging procedure as described in \cite{Andringa:2010it} leads to a geometrical reformulation of non-relativistic gravity called Newton-Cartan gravity \cite{Cartan}. In this reformulation the  trajectory of a particle is described by a geodesic
in a curved so-called Newton-Cartan spacetime. Such a spacetime is described by a (non-invertable) temporal metric $\tau_{\mu\nu}$
and spatial metric $h^{\mu\nu}$, which both are covariantly constant. Via projective relations one can also define the ``inverses'' $\tau^{\mu\nu}$ and $h_{\mu\nu}$ of these metrics.
The equations of motion are defined in terms of the (singular) metric and Christoffel symbols of the Newton-Cartan spacetime. A noteworthy feature is that metric compatibility does not define the Christoffel symbols {\sl uniquely} in terms of (derivatives of) the temporal and spatial metric.
To make contact with a Galilean observer one imposes a set of gauge-fixing conditions which restrict the symmetries to the
acceleration-extended Galilei ones.
The expected equations of motion in terms of a gravitational potential $\Phi(x)$ then follow. The (derivative of the) gravitational potential emerges as the space-time-time component of the Christoffel symbol.

It is natural to extend the above ideas from particles to strings. This will give us information about
the gravitational forces as experienced by a non-relativistic string instead of a particle.
Although the symmetries involved are different, the ideas
are the same as in the particle case discussed above. The starting point in this case is a string moving in a flat Minkowski background.
Taking the non-relativistic limit leads to the action for a non-relativistic string \cite{Gomis:2000bd,Danielsson:2000gi,Gomis:2004pw} that is invariant
under a ``stringy'' version of the Galilei symmetries. The transformations involved, which will be specified later, are
similar to the particle case except that now not only time but also the spatial direction along the string plays a special role.
This leads to an $M_{1,1}$-foliation of spacetime.
Again, the Lagrangian is only invariant up to a total derivative (in the world-sheet coordinates) and hence we obtain an extension of the
``stringy'' Galilei algebra which involves two additional generators $Z_a$ and $Z_{ab}=-Z_{ba}\ (a=0,1)$.\,\footnote{Our notation and conventions can be found in appendix \ref{appendixnotation}.} Due to the extra index structure these
generators provide general extensions rather than central extensions of the stringy Galilei algebra \cite{Brugues:2004an}.

Any two free-falling frames are connected by a stringy Galilei transformation. A ``stringy'' Galilean observer is now defined as an observer with respect to any frame that is accelerated,
  with arbitrary (time and longitudinal coordinate dependent) acceleration,  with respect to a free-falling frame. The corresponding acceleration-extended ``stringy'' Galilei symmetries are obtained by gauging the
translations in the spatial directions transverse to the string by promoting the corresponding
parameters to arbitrary functions of the world-sheet coordinates. These transformations involve the constant transverse translations
and the stringy boost transformations, which are linear in the world-sheet coordinates.

Again, there are two ways to obtain the equations of motion for a stringy Galilean observer. Either we start from the string in a Minkowski background and gauge the transverse translations. In the string case this requires the introduction of a background gravitational potential $\Phi_{\alpha\beta}(x)=\Phi_{\beta\alpha}(x)\ (\alpha=0,1)$, as was also pointed out in \cite{Bagchi:2009my}. This is a striking difference with general relativity where, independent of
whether particles or strings are the basic objects, one always ends up with the same metric function $g_{\mu\nu}(x)$. This is related to the fact that in the non-relativistic case spacetime is a foliation and that the dimension of the foliation space depends on the nature of the
basic object (particles or strings). Alternatively, one gauges the full deformed stringy Galilei algebra and imposes a set of kinematical constraints, like in the particle case. The equation  of motion for $\Phi_{\alpha\beta}(x)$ can be obtained by requiring that it is of second order in the transverse spatial derivatives and invariant under the acceleration-extended stringy Galilei transformations.
In the string case one requires that both the curvature of spatial rotations transverse to the string as well as the curvature of rotations among the foliation directions vanishes. This leads to a flat foliation corresponding to an $M_{1,1}$-foliation
 of spacetime as well as to flat transverse directions. One next introduces the equations of motion making use of the (non-invertable)
 temporal and spatial metric and Christoffel symbols corresponding to the stringy Newton-Cartan spacetime.
To make contact with a stringy Galilean observer one  imposes
 gauge-fixing conditions which reduce  the symmetries to the acceleration-extended stringy Galilei ones.  As expected, the two approaches lead to precisely the same expression for the equation of motion of a fundamental string as well as of the gravitational potential
 $\Phi_{\alpha\beta}(x)$ itself. The (derivative of the) latter
 emerges as a transverse-longitudinal-longitudinal component of the Christoffel symbol.

In order to study  applications of the AdS/CFT correspondence based on the symmetry algebra corresponding to a non-relativistic string
it is necessary to include a (negative) cosmological constant $\Lambda$. It is instructive to first discuss the particle case. In the
relativistic case this means that the Poincar\'e algebra gets replaced by an Anti-de Sitter (AdS) algebra corresponding to a particle
moving in an $\text{AdS}$  background. It is well-known that  one cannot obtain general relativity with a (negative) cosmological constant by gauging the AdS algebra in the same way that one can obtain general relativity by gauging the Poincar\'e algebra. The (technical) reason for this is that one cannot
find a set of (so-called conventional) curvature constraints whose effect is to convert the translation transformations into
general coordinate transformations and, at the same time, to make certain gauge fields to be dependent on others, see e.g. \cite{Ortin:2004ms}. However, we are lucky.
It turns out that, when taking the non-relativistic limit of a particle moving in an AdS background, which is a $\Lambda$-deformation
of the Minkowski background, one ends up with a non-relativitic particle action which is a particular case of the non-relativistic
particle action for a Galilean observer with zero cosmological constant but with the following non-zero-value of the
gravitational potential:
\begin{equation}
\Phi(x^i) = -\tfrac{1}{2}\Lambda x^i x^j \delta_{ij}\,,
\end{equation}
where $\{x^i\}$ are the transverse coordinates. The action is invariant under the so-called
Newton-Hooke symmetries which are a $\Lambda$-deformation of the Galilei symmetries. All Newton-Hooke symmetries can be viewed
as particular time-dependent transverse translations. Therefore, when gauging the transverse translations, it does not
matter whether one gauges the Galilei or Newton-Hooke symmetries, in both cases  one ends up with the same theory but with a different interpretation of the potential. When gauging the Galilei symmetries one interprets the potential $\Phi(x)$ as a purely gravitational potential $\phi(x)$,
i.e.~$\Phi (x)= \phi(x)$. On the other hand, when gauging the Newton-Hooke symmetries one writes $\Phi(x)$ as the sum
of a purely gravitational potential $\phi(x)$ and a $\Lambda$-dependent part, i.e.
\begin{equation}
\Phi (x) = \phi (x) - \tfrac{1}{2}\Lambda x^i x^j \delta_{ij}\,.
\end{equation}
In both cases, turning off gravity amounts to setting $\phi (x) =0$. For $\Lambda=0$ this implies $\Phi (x) =0$ but for $\Lambda \ne 0$ this implies
$\Phi (x^i)= \tfrac{1}{2}\Lambda x^i x^j \delta_{ij}$. These different conditions lead to different surviving symmetries: (centrally extended) Galilei symmetries for $\Lambda=0$ versus (centrally extended) Newton-Hooke symmetries
  \cite{Bacry:1968zf,Gibbons:2003rv}
for $\Lambda\ne 0$.

It is now a relatively straightforward task to generalize the above discussion  to a string moving in an AdS background. Taking the non-relativistic limit
of a string moving in such a background leads to a non-relativistic action that is invariant under a stringy version
of the Newton-Hooke symmetries \cite{Gomis:2005pg,Brugues:2006yd}. Note that this action is $\Lambda$-deformed in two ways: (i) there is a  $\Lambda$-dependent potential
term in the action like in the particle case and (ii) the foliation metric is deformed from $M_{1,1}\ (\Lambda=0)$ to
$\text{AdS}_2\ (\Lambda \ne 0)$. The latter deformation, which leads to an ${\rm AdS}_2$-foliation of spacetime, is trivial in the particle case. All stringy Newton-Hooke symmetries can be viewed as  particular world-sheet dependent transverse translations. It is therefore sufficient to gauge the symmetries for
the case $\Lambda=0$ only, which amounts to gauging the stringy Galilei symmetries.  In a second stage one obtains the $\Lambda\ne 0$ case by a different interpretation of the potential $\Phi_{\alpha\beta}(x)$ and by replacing the flat foliation space by an $\text{AdS}_2$ spacetime.
 To be concrete, in analogy to the particle case, we gauge the  stringy Galilei symmetries only and, next,
write the background potential $\Phi_{\alpha\beta}(x)$, which is needed for this gauging, as
\begin{equation}
\Phi_{\alpha\beta}(x) = \phi_{\alpha\beta}(x) + \tfrac{1}{4}\Lambda\, x^i x^j\, \delta_{ij} \tau_{\alpha\beta}\,,
\end{equation}
where $\phi_{\alpha\beta}(x)$ is the purely gravitational potential and
$\tau_{\alpha\beta}$ is an $\text{AdS}_2$-metric. At the same time we have replaced the flat foliation by an $\text{AdS}_2$ space
leading to an $\text{AdS}_2$-foliation of spacetime.\,\footnote{When gauging the full (deformed) stringy Galilei symmetries one of the
kinematical constraints which have to be imposed in order to restrict to a stringy Galilean observer, for $\Lambda \ne 0$,  is that the
curvature corresponding to rotations amongst the longitudinal directions  is proportional to $\Lambda$. This leads to a flat foliation for $\Lambda=0$ but an $\text{AdS}_2$-foliation for $\Lambda\ne 0$.}

In this way it is a relatively simple manner to obtain the geodesic equations of motion for a fundamental string in a cosmological background and to derive the equations
of motion for the potential $\Phi_{\alpha\beta}(x)$ itself. We will give the explicit expressions in the second part of this paper.

This work is organized as follows. In Section 2 we review, as a warming-up exercise, the particle case for zero cosmological constant.
The gauging of the Bargmann algebra, i.e.~the centrally extended Galilei algebra,  will only be discussed at the level of the symmetries; for full details we refer to
\cite{Andringa:2010it}.
In Section 3 we derive the relevant expressions for the string case.
In particular, we discuss the gauging of the full (deformed) stringy Galilei symmetries.
The extension to a non-zero cosmological constant will be discussed in Section 4 using the observations mentioned above.
In this section we will present explicit expressions for the equation of motion for a non-relativistic fundamental string  in a cosmological background and the equations of motion for
the potential $\Phi_{\alpha\beta}(x)$. These two equations together describe the dynamics of ``stringy'' Newton-Cartan gravity
as observed by a ``stringy'' Galilean observer.
The potential application of this theory to the AdS/CFT correspondence based on the non-relativistic  Newton-Hooke
algebra will be briefly discussed in the Conclusions.

\section{The Particle Case}
\label{The Particle Case}
Our starting point is the action describing a particle of mass $m$ moving in a $D$-dimensional Minkowski spacetime, i.e. $\Lambda=0$,
with metric $\eta_{\mu\nu}\ (\mu = 0,1,\cdots ,D-1)$:
\be
S =  - m\int d\tau\, \sqrt{-\eta_{\mu\nu}\dot{x}^{\mu}\dot{x}^{\nu}} \,. \label{RelPointPart1}
\ee
 Here $\tau$ is the evolution parameter parametrizing the worldline and the dot indicates differentiation
 with respect to $\tau$. We have taken the speed of light to be $c=1$. This action is invariant under worldline reparametrizations. The Lagrangian,
 defined by $S = \int L d\tau$, is invariant under the Poincar\'e transformations
 with parameters $\lambda^\mu{}_\nu$ (Lorentz transformations) and $\zeta^\mu$ (translations):
 \be\label{Poincare}
\delta x^{\mu} = \lambda^{\mu}{}_{\nu}x^{\nu} + \zeta^{\mu} \,.
\ee

Following  \cite{Gomis:2004pw,Gomis:2005pg} we  take the non-relativistic limit by rescaling the longitudinal coordinate $x^0\equiv t$ and the mass $m$
with a parameter $\omega$ and taking  $\omega >> 1$:
\be
x^0 \rightarrow \omega x^0, \ \ \ m \rightarrow \omega m, \ \ \ \omega >> 1 \,. \label{particlex0rescaling}
\ee
This rescaling is such that the kinetic term remains finite. This results into the following action:
\be
S \approx
- m\omega^2 \int \dot{x}^0 \Bigl(1 - \frac{\dot{x}^i \dot{x}_i}{2\omega^2 (\dot{x}^0)^2}\Bigr) d\tau \,,\hskip .5truecm
i=1,\cdots ,D-1\,.
\label{limitparticleaction1}
\ee
The first term on the right-hand-side, which is a total derivative, can be cancelled by coupling the particle to a
constant background gauge field $A_{\mu}$
by adding a term
\be
S_I = m \int A_{\mu}\dot{x}^{\mu} d\tau  \,, \label{gaugebackgroundfield}
\ee
and choosing $A_0 = \omega^2$ and $A_i = 0$  \cite{Gomis:2000bd}. Because this $A_{\mu}$ can be written as a total derivative the associated field-strength vanishes, such that no dynamics for the background gauge field is introduced. The limit $\omega \rightarrow \infty$ then yields the
following non-relativistic action
\be
S = \frac{m}{2}\int \frac{\dot{x}^i \dot{x}^j\delta_{ij}}{\dot{x}^0}d\tau \,. \label{nrparticlereparaminvaction}
\ee
This action is invariant under worldline reparametrizations and the following Galilei symmetries
\be
\delta x^0 = \zeta^0, \ \ \ \ \delta x^i = \lambda^i{}_j x^j + v^i x^0 + \zeta^i \,, \label{Galileisymmetries}
\ee
where $\big(\zeta^0\,,\zeta^i\,,\lambda^i{}_j\,, v^i\big)$ parametrize a (constant) time translation, space translation, spatial rotation and boost
transformation, respectively. The equations of motion corresponding to the action \eqref{nrparticlereparaminvaction} are\,\footnote{One can check that the equation of motion for $\{x^0\}$ and $\{x^i\}$
corresponding to the action  \eqref{nrparticlereparaminvaction} are not independent; the first can be derived from the latter. When we will include gravity in \eqref{nrparticlereparaminvaction} via the worldline-reparametrization invariant coupling $\dot{x}^0\Phi(x)$, see eq.\eqref{actionparticle}, this will again be the case.}
\be
\ddot{x}^i = \frac{\ddot{x}^0}{\dot{x}^0} \dot{x}^i \label{EOMparticleflat} \,.
\ee
It turns out that the non-relativistic Lagrangian \eqref{nrparticlereparaminvaction}  is invariant under boosts only up to a total $\tau$-derivative, i.e.,
\be
\delta L = \frac{d}{d\tau}(mx^i v^j\,\delta_{ij})\,. \label{particleboostquasiinvariance}
\ee
This leads to a modified Noether charge giving rise to a centrally extended Galilei algebra containing an extra
so-called central charge generator $Z$, see e.g.~\cite{levyleblond69,Marmo:1987rv}. This centrally extended Galilei algebra is called the Bargmann algebra
\cite{Bargmann}.

The above results apply to free-falling frames without any gravitational interactions.
Such frames are connected to each other via the Galilei symmetries \eqref{Galileisymmetries}.
We now wish to extend these results to include  frames that apply to a Galilean observer, i.e.~that are accelerated with respect to the free-falling frames,
with arbitrary (time-dependent) acceleration. As explained in the introduction we can do this via two distinct gauging procedures.
The first procedure is convenient if one is only interested in the physics experienced by a Galilean observer. In
that case it is sufficient to gauge the  transverse translations by replacing the constant parameters $\zeta^i$ by arbitrary time-dependent functions $\zeta^i\  \rightarrow \  \xi^i(x^0)$. Applying this gauging to the action \eqref{nrparticlereparaminvaction}
leads to the following gauged action containing the gravitational potential $\Phi(x)$:\,\footnote{Note that $\Phi(x)$ is a background field representing a set of coupling constants from the world-line point of view. Since these coupling constants also transform we are dealing
 not with a ``proper'' symmetry but with a ``pseudo'' or ``sigma-model'' symmetry, see, e.g.~\cite{Moore:1984dc,Hull:1995gk}.}
\be
S = \frac{m}{2}\int d\tau\, \Bigl( \frac{\dot{x}^i \dot{x}^j\delta_{ij}}{\dot{x}^0}\, - 2\dot{x}^0\Phi(x) \Bigr) \,.  \label{actionparticle}
\ee

 The action \eqref{actionparticle} is invariant under worldline reparametrizations and the acceleration-extended symmetries (we write $x^0$ as $t$ from now on)
\be
\delta t = \zeta^0, \ \ \ \ \delta x^i = \lambda^i{}_j x^j + \xi^i(t)\,,  \label{Galilei1}
\ee
provided that the ``background gauge field'' $\Phi(x)$ transforms as follows:
\be
\delta\Phi(x) = - \frac{1}{\dot{t}} \frac{d}{d\tau} \Bigl(\frac{\dot{\xi}_i}{\dot{t}}\Bigr)x^i  + \p_0 g(t)\,.\label{Galilei2}
\ee
The second term with the arbitrary function $g(t)$ represents a standard ambiguity in any potential describing a force and gives a boundary term in the action \eqref{actionparticle}.
This action leads to the following modified equation of motion describing a particle moving
in a gravitational potential:
\be
\ddot{x}^i + (\dot{t})^2 \delta^{ij}\p_j \Phi(x) = \frac{\ddot{t}}{\dot{t}}\, \dot{x}^i \,.\label{first1}
\ee
Notice how \eqref{Galilei2} and \eqref{first1} simplify if one takes the static gauge
\begin{equation}\label{particlestaticgauge}
t=\tau\,,
\end{equation}
for which $\dot{t}=1$ and $\ddot{t}=0$. Using this static gauge we see that
for {\sl constant} accelerations ${\ddot \xi}^i = \text{constant}$,   it is sufficient to introduce a time-independent potential $\Phi(x^i)$ but that for time-dependent accelerations we need a
potential $\Phi(x)$ that depends on both the time and the transverse spatial directions.\\

The equation of motion of $\Phi(x)$ itself is easiest obtained by requiring that it is second order in spatial derivatives and invariant
under the acceleration-extended Galilei symmetries \eqref{Galilei1} and \eqref{Galilei2}. Since the variation of $\Phi(x)$,
see eq.~\eqref{Galilei2}, contains an arbitrary function of time and is linear in the transverse coordinate, it is clear that
the unique second-order differential operator satisfying this requirement is the Laplacian $\Delta \equiv \delta^{ij}\partial_i\partial_j$.
Requiring that the source term is provided by the mass density function $\rho (x)$, which transforms as a scalar with respect to \eqref{Galilei1}, this leads to the following Poisson equation
\be
\triangle \Phi (x) = V_{D-2} G \rho (x)\,,\label{first2}
\ee
where we have introduced Newton's constant $G$ for dimensional reasons, and $V_{D-2}$ is the volume of a $(D-2)$-dimensional sphere.

The second gauging procedure is relevant if one is interested in describing the physics in more frames than the set of
accelerated ones. In that case one needs to gauge {\sl all} the symmetries of the Bargmann algebra. This gauging has been
described  in \cite{Andringa:2010it}. We will not repeat the full procedure here but explain the basic points and  concentrate on the symmetries involved.
The starting point is the Bargmann algebra which consists of time and space translations, spatial rotations, boosts and central charge transformations. In the gauging procedure one associates a gauge field to each of the symmetries (for our index-notation, see appendix A):
\begin{eqnarray}
\tau_\mu&:&\ \ \ \text{time translations}\nonumber \\[.1truecm]
e_\mu{}^{a'}&:&\ \ \ \text{space translations}\nonumber \\[.1truecm]
\omega_\mu{}^{a'\underline{0}}&:&\ \ \ \text{boosts} \\[.1truecm]
\omega_\mu{}^{a'b'}&:&\ \ \ \text{spatial rotations}\nonumber \\[.1truecm]
m_\mu&:&\ \ \ \text{central charge transformations}\,.\nonumber
\end{eqnarray}
Furthermore, the constant parameters describing the transformations are promoted to arbitrary functions of the spacetime coordinates $\{x^\mu\}$:
\begin{eqnarray}
\tau(x^\mu)&:&\ \ \ \text{time translations}\nonumber \\[.1truecm]
\zeta^{a'}(x^\mu)&:&\ \ \ \text{space translations}\nonumber \\[.1truecm]
\lambda^{a'\underline{0}}(x^\mu)&:&\ \ \ \text{boosts} \\[.1truecm]
\lambda^{a'b'}(x^\mu)&:&\ \ \ \text{spatial rotations}\nonumber \\[.1truecm]
\sigma(x^\mu)&:&\ \ \ \text{central charge transformations}\,.\nonumber\label{parameters}
\end{eqnarray}
Besides these transformations all gauge fields transform under general coordinate transformations
with parameters $\xi^\mu(x^\mu)= (\xi^0(x^\mu)\,, \xi^i(x^\mu))$. As a first step in the gauging procedure
one imposes a set of so-called conventional constraints on the curvatures of the gauge fields. The purpose of these constraints
is two-fold. First of all, it has the effect that the time and space translations become equivalent to general coordinate transformations
modulo the other symmetries of the algebra \cite{book}. This can be seen from the following identity, which relates the general coordinate transformation of a gauge field $B_{\mu}{}^A$ to its curvature $R_{\mu\lambda}{}^A$ and the other gauge transformations in the theory with field-dependent parameters:
\begin{equation}
   \delta_{gct}(\xi^{\lambda})B_{\mu}{}^A +
          \xi^\lambda R_{\mu\lambda}{}^A
   - \sum_{\substack{\{C\}}} \delta(\xi^{\lambda}B_{\lambda}{}^{C})B_{\mu}{}^A = 0 \,.
\label{removingHandP}
\end{equation}
Secondly, the conventional constraints enable one to solve for the
gauge fields $\omega_\mu{}^{a'\underline{0}}$ and $\omega_\mu{}^{a'b'}$  in terms of the other ones \cite{Andringa:2010it}:
\begin{eqnarray}
\omega_{\mu}{}^{a'b'} &=& 2e^{\rho [a'}\p_{[\rho}e_{\mu]}{}^{b']} - e^{\rho a'} e^{\nu b'}e_{\mu}{}^{c'} \p_{[\nu}e_{\rho]}{}^{c'} - \tau_{\mu}e^{\rho [a'}\omega_{\rho}{}^{b']\underline{0}} \,, \label{omegaij} \\ [.15truecm]
\omega_{\mu}{}^{a'\underline{0}} & =& e^{\nu a'}\p_{[\mu}m_{\nu]} + e^{\nu a'}\tau^{\rho} e_{\mu}{}^{b'} \p_{[\nu}e_{\rho ]}{}^{b'} + \tau_{\mu}\tau^{\nu} e^{\rho a'}\p_{[\nu}m_{\rho ]} + \tau^{\nu} \p_{[\mu}e_{\nu ]}{}^{a'} \label{omegai0} \,.
\end{eqnarray}
The same constraints have a third effect, namely that the gauge field $\tau_\mu$ of time translations can be written as the spacetime derivative
of an arbitrary  function $f(x)$:
\be
\tau_\mu = \partial_\mu f(x)\,.\label{arbf}
\ee
At this point the symmetries of the theory are the general coordinate transformations plus the
boosts, spatial rotations and central charge transformations, all with  parameters that are arbitrary functions of the spacetime coordinates.

The gauge fields $\tau_\mu$ and $e_\mu{}^{a'}$ of  time and spatial translations are identified as the (singular) temporal and spatial vielbeins. One
may also introduce their inverses (with respect to the temporal and spatial subspaces) $\tau^\mu$ and $e^\mu{}_{a'}$:
\begin{eqnarray}
e_{\mu}{}^{a'} e^{\mu}{}_{b'} & = &\delta_{b'}^{a'}, \ \ \ \ \ e_{\mu}{}^{a'} e^{\nu}{}_{a'}  = \delta^{\nu}_{\mu} -
\tau_{\mu}\tau^{\nu}\,,\ \ \ \ \ \tau^{\mu}\tau_{\mu} = 1\,,\nonumber\\[.15truecm]
\tau^{\mu} e_{\mu}{}^{a'} & = &0, \ \ \ \ \ \ \ \tau_{\mu}e^{\mu}{}_{a'} =  0\,. \label{NCvielbeinrelations}
\end{eqnarray}
The spatial and temporal vielbeins  define
spatial and temporal metrics  as follows:
\begin{eqnarray}
\tau_{\mu\nu} &=& \tau_\mu\tau_\nu\,, \hskip 2.3truecm \tau^{\mu\nu} = \tau^\mu\tau^\nu\,,\nonumber \\[.15truecm]
h_{\mu\nu} &=& e_\mu{}^{a'}e_\nu{}^{b'}\,\delta_{a'b'}\,,\hskip 1.5truecm
h^{\mu\nu} = e^\mu{}_{a'}e^\nu{}_{b'}\,\delta^{a'b'}\,. \label{metricvielbeinrelation}
\end{eqnarray}
A $\Gamma$-connection can be introduced by
assuming the vielbein postulates:
\begin{eqnarray}
 \partial_{\mu}e_{\nu}{}^{a'} - \omega_{\mu}{}^{a'b'}e_{\nu}{}^{b'}
                             - \omega_{\mu}{}^{a'\underline{0}}\tau_{\nu} -
     \Gamma_{\nu\mu}^{\rho}e_{\rho}{}^{a'} = 0\,,
\hskip 1truecm
 \partial_\mu \tau_\nu -
  \Gamma^\lambda_{\nu\mu}\tau_\lambda = 0\,.
\label{Hpost}
\end{eqnarray}
These vielbein postulates state that $\tau_\mu$ is covariantly constant whereas $e_{\mu}{}^{a'}$ is not,\footnote{Remember that $\na_{\rho}h^{\mu\nu}=0$ and $\na_{\rho}h_{\mu\nu}\neq 0$.} and can be uniquely solved for the $\Gamma$-connection, giving
\begin{align}
\Gamma^{\rho}_{\nu\mu} & = \tau^{\rho}\partial_{(\mu}\tau_{\nu)}
 + e^{\rho}{}_{a'} \Bigl(	\partial_{(\mu}e_{\nu)}{}^{a'} - \omega_{(\mu}{}^{a'b'}e_{\nu)}{}^{b'}  - \omega_{(\mu}{}^{a'\underline{0}}\tau_{\nu)}\Bigr) \,, \label{NHGamma}
\end{align}
where the dependent fields $\omega_{\mu}{}^{a'b'}$ and $\omega_{\mu}{}^{a'\underline{0}}$ are given by \eqref{omegaij} and \eqref{omegai0}. If we plug in these explicit solutions, one obtains
\begin{align}
\Gamma^{\rho}_{\nu\mu} & = \tau^{\rho}\partial_{(\mu}\tau_{\nu)} + \frac{1}{2}h^{\rho\sigma} \Bigl(\p_{\nu}h_{\sigma\mu} + \p_{\mu}h_{\sigma\nu} - \p_{\sigma}h_{\mu\nu}\Bigr)\nonumber + h^{\rho\sigma}K_{\sigma(\mu} \, \tau_{\nu)} \,, \nonumber\\
K_{\mu\nu} & = 2 \p_{[\mu}m_{\nu]} \,. \label{NCparticleConhtauK}
\end{align}
The Riemann tensor can be obtained, using the vielbein postulates, from the curvatures of the spin connection fields:
\begin{equation}
R^{\mu}_{\ \nu\rho\sigma}(\Gamma)  = - e^{\mu}{}_{a'}R_{\rho\sigma}{}^{a'b'}(M'')e_{\nu b'} - e^{\mu}{}_{a'}R_{\rho\sigma}{}^{a'\underline{0}}(M') \tau_{\nu} \,. \label{NCRiemanntensor}
\end{equation}
At this stage the independent gauge fields are given by $\{\tau_{\mu}, e_{\mu}{}^{a'}, m_{\mu}\}$. The dynamics of the Newton-Cartan point particle is now described by the following action \cite{Kuchar:1980tw}:
\be
L = \frac{m}{2} \Bigl(\frac{h_{\mu\nu}\dot{x}^{\mu} \dot{x}^{\nu}}{\tau_{\rho}\dot{x}^{\rho}} - 2m_{\mu}\dot{x}^{\mu}\Bigr) \,. \label{altparticleaction}
\ee
Alternatively, this action can be written as
\begin{equation}\label{alternative}
L = \frac{m}{2}N^{-1}\dot{x}^{\mu}\dot{x}^{\nu}\Bigl(h_{\mu\nu} - 2m_{\mu}\tau_{\nu}\Bigr)
\end{equation}
with $N \equiv \tau_{\mu}\dot{x}^{\mu}$.

The first term in this Lagrangian can be seen as the covariantization of the Lagrangian of \eqref{nrparticlereparaminvaction} with the Newton-Cartan metrics $h_{\mu\nu}$ and $\tau_{\mu}$. The presence of the central charge gauge field  $m_\mu$ represents the ambiguity when trying to solve the $\Gamma$-connection in terms of the (singular) metrics of Newton-Cartan spacetime. The Lagrangian \eqref{altparticleaction} is quasi-invariant under the gauged Bargmann algebra; under $Z$-transformations $\delta m_{\mu}=\p_{\mu}\sigma$ the Lagrangian \eqref{altparticleaction} transforms as a total derivative, while for the other transformations the Lagrangian is invariant. In fact, the
 $m_{\mu}\dot{x}^{\mu}$ term  in \eqref{altparticleaction} is needed in order to render the action invariant under boost transformations which transform both the spatial metric $h_{\mu\nu}$ and the central charge gauge field $m_\mu$ as follows:
\be
\delta h_{\mu\nu}=2\lambda^{a'}{}_{\underline{0}} e_{(\mu}{}^{a'} \tau_{\nu )}\,,\hskip 1truecm \delta m_{\mu} = \lambda^{a'}{}_{\underline{0}}e_{\mu}{}^{a'} \,.
\ee
Varying the Lagrangian \eqref{altparticleaction} gives, after a lengthy calculation,\footnote{Some details are given in appendix \ref{NCgeodesicdetails}.} the geodesic equation
\be
{\ddot x}^\mu +\Gamma^\mu_{\nu\rho}{\dot x}^\nu {\dot x}^\rho = \frac{\dot{N}}{N}\dot{x}^{\mu} \,. \label{geodesiceq}
\ee
Here $N \equiv \tau_{\mu}\dot{x}^{\mu} = \dot{f}$, which in adapted coordinates becomes $N=\dot{t}$, and the $\Gamma$-connection is given by \eqref{NHGamma}. The geodesic equation \eqref{geodesiceq} can be regarded as the covariantization of \eqref{first1}.

Unlike the particle dynamics, the gravitational dynamics cannot be obtained from an action in a straightforward way, see e.g. \cite{Duval:1984cj}. The equation describing the dynamics of Newton-Cartan spacetime may be written in terms of the Ricci-tensor
of the $\Gamma$-connection as follows:
\be
 R_{\mu\nu}(\Gamma) = V_{D-2} G \rho \tau_{\mu\nu}\,.  \label{Einsteineq}
\ee
To make contact with the equations for a Galilean observer, derived in the first gauging procedure, one must
impose the kinematical constraint that the curvature corresponding to the
$(D-1)$-dimensional spatial rotations equals to zero:
\be
R_{\mu\nu}{}^{a^\prime b^\prime}(M'') = 0\,.\label{kinc}
\ee
Here $M''$ refers to the generators of spatial rotations. It should be stressed that one is not forced to impose this curvature constraint, and one could stay more general and try to solve the resulting theory of gravity for a curved transverse space. We will see that the constraint \eqref{kinc} can be considered as an ansatz for the  transverse Newton-Cartan metric $h^{\mu\nu}$ to be flat.
 It is also convenient to choose so-called {\sl adapted} coordinates in which the function $f(x)$ in eq.~\eqref{arbf} is set equal to the time or foliation coordinate $t:
f(x)=t$. This reduces the general coordinate transformations to constant time translations and
 spatial translations with an arbitrary space-time dependent parameter.

The kinematical constraint \eqref{kinc}
enables us to do two things. First, we can now choose a flat Cartesian coordinate system in the $(D-1)$
spatial dimensions, because the transverse space is flat as can be seen from eq.~\eqref{NCRiemanntensor}:\footnote{Note that eq.~\eqref{vanishingRiemannduetoRJ} already follows from the equations of motion \eqref{Einsteineq} in the case of $D=4$, because in three dimensions a vanishing Ricci tensor implies a vanishing Riemann tensor.}
\be
R^i{}_{jkl}(\Gamma) = 0 \,. \label{vanishingRiemannduetoRJ}
\ee
The solution \eqref{omegaij} implies that the spatial components $\omega_i{}^{a^\prime b^\prime}$ of the gauge field of spatial rotations
is zero in such a coordinate system, which expresses the fact that the transverse Christoffel symbols vanish:
\be
\Gamma^i_{jk} \sim \delta^i_{a'} \delta_{b'}^{j}\, \omega_k {}^{a^\prime b^\prime} = 0 \,. \label{Gammaikj=0}
\ee
This choice of coordinates
restricts the spatial rotations to those that have a time-dependent parameter only.
Second, due to the same
kinematical constraint \eqref{kinc} the time component  $\omega_0{}^{a^\prime b^\prime}$ of the same gauge field is a pure gauge; $R_{\mu\nu}{}^{a^\prime b^\prime}(M'')$ is the field-strength of an $SO(D-1)$ gauge theory and contains only $\omega_{\mu}{}^{a^\prime b^\prime}$, as can be seen from \eqref{NHcurvatures}. As such the constraint \eqref{kinc} allows one to gauge-fix $\omega_{\mu}{}^{a^\prime b^\prime}$ to zero,\footnote{Explicitly, one can write $R_{\mu\nu}{}^{a^\prime b^\prime}(M'') = 2 D_{[\mu}\omega_{\nu]}{}^{a'b'}$ and $\delta \omega_{\mu}{}^{a'b'} = D_{\mu}\lambda^{a'b'}$, where $D_{\mu}$ is the gauge covariant derivative. Putting $R_{\mu\nu}{}^{a^\prime b^\prime}(M'') = 0$ imposes the constraint $\omega_{\mu}{}^{a'b'} = D_{\mu}f^{a'b'}$ on the gauge field for some $f^{a'b'}$. Performing then a gauge transformation on $\omega_{\mu}{}^{a^\prime b^\prime}$ and choosing the gauge parameter to be $\lambda^{a'b'} = -f^{a'b'}$, the result follows.} and this restricts the spatial rotations
to having constant parameters only.  Via \eqref{NHGamma} one can show that this implies
\be
\Gamma^i_{0j} \sim \delta^i_{a'} \delta_{b'}^{j}\, \omega_0{}^{a^\prime b^\prime} = 0 \,. \label{Gammai0j=0}
\ee
The same choice of a Cartesian coordinate system also restricts the spatial translations to having only
time-dependent parameters. This reduces the symmetries acting on the spacetime coordinates to the acceleration-extended
Galilei symmetries given in eq.~\eqref{Galilei1}. The central charge transformations now only depend on time and do not act on the spacetime coordinates. The vielbein postulate  tells us that the only remaining connection component $\Gamma^i_{00}$ can be written as $\Gamma^i_{00} = \p^i \Phi(x)$, where
\be
\Phi (x)= m_0(x) - \tfrac{1}{2}\delta_{ij}\tau^i (x)\tau^j(x) + \p_0 m(x)  \,. \label{gravpot}
\ee
Here $m_0$ and $\p_i m$ are the time component and spatial gradient components of the extension gauge field $m_\mu$, and $\tau^i$ are the space components of the inverse temporal vielbein $\tau^\mu$. Using the transformation properties of $\Gamma^i_{00}$ one can show that $\Phi(x)$, defined by eq.~\eqref{gravpot}, indeed transforms like in eq.~\eqref{Galilei2} under the acceleration-extended Galilei symmetries.\,\footnote{The fact that $\Phi$ transforms with the double time derivative
of $\xi^i$ shows that it indeed transforms as a component of the $\Gamma$-connection.}

One can show that after gauge-fixing the Newton-Cartan symmetries to the accele\-ration-extended Galilei symmetries, as described above,  the Lagrangian \eqref{altparticleaction} reduces to
\be
L = \frac{m}{2} \Bigl( \frac{\delta_{ij}\dot{x}^{i} \dot{x}^{j}}{\dot{x}^0} + \dot{x}^0 (\delta_{ij}\tau^i \tau^j	- 2m_0 - 2\p_0 m)	\Bigr) \,,
\ee
where a boundary term has been discarded.\footnote{We have made use of the fact that, because $x^{\mu}=x^{\mu}(\tau)$, the $\tau$-derivative of a general function $f(x)$ can be written as $\dot{f}(x) = \dot{x}^0 \p_0 f(x) + \dot{x}^i \p_i f(x)$, which in the static gauge becomes $\dot{f}(x) =  \p_0 f(x) + \dot{x}^i \p_i f(x)$.} Upon comparison with the action \eqref{actionparticle} this again identifies the potential
as in \eqref{gravpot}. Note that the $\tau^i \dot{x}^i$ terms cancel, reflecting the choice of gauge \eqref{Gammai0j=0} and indicating that this particular reference frame is non-rotating. Similarly, eq.\eqref{Einsteineq} reduces in this gauge to the Poisson equation \eqref{first2}.

As expected, having the same symmetries, the equations of motion
 \eqref{geodesiceq} and \eqref{Einsteineq} reduce to precisely the  equations of motion \eqref{first1} and \eqref{first2}
 we obtained in the  first gauging procedure. \\

\section{From Particles to Strings}

We now consider instead of  particles of mass $m$ strings with tension $T$ moving in a $D$-dimensional
Minkowski spacetime, with metric $\eta_{\mu\nu}\ (\mu = 0,1,\cdots ,D-1)$. The action describing the dynamics of such a string is given by (we take $c=1$)
\begin{equation}
S = -T\int d^2 \sigma \sqrt{-\gamma}\,, \label{RelstringPart1}
\end{equation}
where $\sigma^{{\bar\alpha}}\ ({\bar\alpha}=0,1)$ are the world-sheet coordinates and $\gamma$ is the determinant of the induced
world-sheet metric $\gamma_{\bar\alpha\bar\beta}$:
\be
\gamma_{\bar{\alpha}\bar{\beta}} = \p_{\bar{\alpha}} x^{\mu}\p_{\bar{\beta}} x^{\nu} \eta_{\mu\nu}\,. \label{relwsmetric}
\ee
The action \eqref{RelstringPart1} is invariant under world-sheet reparametrizations. Like in the particle case,
the Lagrangian corresponding to this action is invariant under Poincar\'{e} transformations in the
target spacetime, see eq.~\eqref{Poincare}.

Following  \cite{Gomis:2004pw,Gomis:2005pg} we  take the non-relativistic limit by rescaling the longitudinal coordinate $x^\alpha = (x^0\equiv t,x^1)$
with a parameter $\omega$ and taking  $\omega >> 1$:\,\footnote{Note that, unlike the particle case, the parameter $T$ does not get rescaled.}
\be
x^\alpha \rightarrow \omega x^\alpha\,,  \ \ \ \omega >> 1 \,. \label{stringxalpharescaling}
\ee
This results into the following action $(i=2,\cdots ,D-1)$:
\be
S \approx - T \omega^2 \int  d^2 \sigma \sqrt{-\bar{\gamma}} \Bigl(1 + \frac{1}{2\omega^2}\bar{\gamma}^{\bar{\alpha}\bar{\beta}}\p_{\bar{\alpha}} x^i \p_{\bar{\beta}} x^j \delta_{ij} \Bigr)\,,\label{followinga}				
\ee
where ${\bar\gamma}_{\bar\alpha\bar\beta}$ is the pull-back of the longitudinal metric $\eta_{\alpha\beta}$, i.e.
\be
\bar{\gamma}_{\bar{\alpha}\bar{\beta}} = \p_{\bar{\alpha}} x^{\alpha}\p_{\bar{\beta}} x^{\beta} \eta_{\alpha\beta} \,. \label{NRpullbacklongmetric}
\ee
Unlike the worldsheet metric \eqref{relwsmetric}, the pull-back used in \eqref{NRpullbacklongmetric} is given by a $2 \times 2$-matrix, and as such is invertible. This means that the inverse metric $\bar{\gamma}^{\bar{\alpha}\bar{\beta}}$ can be explicitly given: it is the pull-back of the longitudinal inverse metric $\eta^{\alpha\beta}$,
\be
\bar{\gamma}^{\bar{\alpha}\bar{\beta}} = \p_{\alpha} \sigma^{\bar{\alpha}} \p_{\beta} \sigma^{\bar{\beta}}  \eta^{\alpha\beta} \,, \label{NRpullbacklongmetricinverse}
\ee
such that $\bar{\gamma}^{\bar{\alpha}\bar{\beta}} \bar{\gamma}_{\bar{\beta}\bar{\e}} = \delta^{\bar{\alpha}}_{\bar{\e}}$. \\

The divergent term on the right hand side of eq.~\eqref{followinga} is a total world-sheet derivative \cite{Gomis:2004pw}. This can be seen by using the identity $\eta_{[\beta[\alpha}\eta_{\gamma]\delta]} = - \tfrac{1}{2}\e_{\beta\delta}\e_{\alpha\gamma}$, which holds in two dimensions and in which $\e_{\alpha\gamma}$ is the two-dimensional epsilon symbol. This allows one to write
\be
\sqrt{-\bar{\gamma}} = \p_{\bar{\alpha}}\Bigl(\tfrac{1}{2} \e^{\bar{\alpha}\bar{\gamma}}\e_{\alpha\gamma} x^{\alpha}\p_{\bar{\gamma}} x^{\gamma}  \Bigr) \,.
\ee
The divergent term can be canceled by coupling the string to a constant background 2-form potential
$B_{\mu\nu}$ via the following Wess-Zumino term:
\be
S_I = T \int  d^2\sigma \e^{\bar{\alpha}\bar{\beta}}\p_{\bar{\alpha}} x^{\mu} \p_{\bar{\beta}} x^{\nu} B_{\mu\nu} \,, \label{gaugebackgroundfield2}
\ee
and choosing the constant field components $B_{\mu\nu}$ such that
\be
 B_{\alpha\beta} = \frac{1}{2} \omega^2 \e_{\alpha\beta}, \ \ \ \ B_{i\alpha} = B_{ij} = 0 \,.
\ee
The resulting field-strength of $B_{\mu\nu}$ is zero, similar to the particle case. The limit $\omega \rightarrow \infty$ of the sum of \eqref{followinga} and \eqref{gaugebackgroundfield2} then leads to the following non-relativistic action:
\be
S = - \frac{T}{2} \int  d^2 \sigma \sqrt{-\bar{\gamma}} \Bigl(\bar{\gamma}^{\bar{\alpha}\bar{\beta}}\p_{\bar{\alpha}} x^i \p_{\bar{\beta}} x^j \delta_{ij} \Bigr) \,.\label{nonrelaction}
\ee
This action is invariant under world-sheet reparametrizations and the following ``stringy'' Galilei symmetries:
\be\label{stringyG}
\delta x^{\alpha} = \lambda^{\alpha}{}_{\beta} x^{\beta} + \zeta^{\alpha}, \ \ \ \ \delta x^i = \lambda^i{}_j x^j + \lambda^i{}_{\beta} x^{\beta} + \zeta^i \,,
\ee
where $(\zeta^\alpha\,, \zeta^i\,, \lambda^i{}_j\,, \lambda^i{}_\alpha\,, \lambda^\alpha{}_\beta)$ parametrize a (constant) longitudinal
translation, transverse translation, transverse rotation, ``stringy'' boost transformation and longitudinal rotation, respectively. As for the point particle, the equations of motion for the longitudinal and transverse components are not independent. The
equations of motion for $\{x^i\}$ corresponding to the action \eqref{nonrelaction} are given by
\begin{align}
\p_{\bar{\alpha}}& \Bigl( \sqrt{-\bar{\gamma}} \bar{\gamma}^{\bar{\alpha}\bar{\beta}} \p_{\bar{\beta}} x^i \Bigr) = 0 \,.
\end{align}
The non-relativistic Lagrangian defined by \eqref{nonrelaction}
is invariant under a stringy boost transformation only up to a total world-sheet divergence:
\be
\delta L = \p_{\bar{\alpha}}\Bigl(-T \sqrt{-\bar{\gamma}} \frac{\p \sigma^{\bar{\alpha}}}{\partial x^{\alpha}} \lambda_i{}^{\alpha} x^i \Bigr) \,, \label{Galstringtotderivative}
\ee
where \eqref{NRpullbacklongmetricinverse} has been used. This leads to a modified Noether charge  giving rise to an extension of the stringy Galilei algebra containing two extra generators: $Z_a$ and
$Z_{ab}\ (a=(0,1))$ \cite{Brugues:2004an}. The corresponding extended stringy Galilei algebra is given in appendix \ref{ExtStringGalalg}.

We now wish to connect to the physics as experienced by a ``stringy'' Galilean observer by gauging  the translations in the spatial directions
transverse to the string. In this procedure we replace the constant parameters $\zeta^i$ by functions $\xi^i(x^\alpha)$ depending only on the longitudinal coordinates. Applying this gauging to the non-relativistic action \eqref{nonrelaction} leads to the
following gauged action containing a gravitational potential
$\Phi_{\alpha\beta}$:
\be
S = - \frac{T}{2} \int  d^2 \sigma \sqrt{-\bar{\gamma}} \, \Bigl(\bar{\gamma}^{\bar{\alpha}\bar{\beta}}\p_{\bar{\alpha}} x^i \p_{\bar{\beta}} x^j \delta_{ij} - 2\eta^{\alpha\beta} \Phi_{\alpha\beta}\Bigr)\,. \label{nonrelstringpotent}
\ee
This action can be compared with the point particle action \eqref{actionparticle}.\,\footnote{Note that $\bar{\gamma}_{\bar{\alpha}\bar{\beta}}$
corresponds to a factor  $-({\dot x}^0)^2$ in the particle action.}
 The string action \eqref{nonrelstringpotent} is invariant under world-sheet reparametrizations and the acceleration-extended stringy Galilei symmetries \cite{Brugues:2004an}
\be\label{sGalilei1}
\delta x^{\alpha} = \lambda^{\alpha}{}_{\beta} x^{\beta} + \zeta^{\alpha}, \ \ \ \ \delta x^i = \lambda^i{}_j x^j  + \xi^i(x^{\alpha}) \,.
\ee
The local transverse translations are only realized provided that the background potentials $\Phi_{\alpha\beta}$ transform as follows:
\be\label{sGalilei2}
\delta \Phi_{\alpha\beta} = - \frac{1}{2\sqrt{-\bar{\gamma}}} \, \eta_{\alpha\beta} \,  \p_{\bar{\alpha}}\Bigl( \sqrt{-\bar{\gamma}} \, \bar{\gamma}^{\bar{\alpha}\bar{\beta}}\,\p_{\bar{\beta}} \xi_i  \Bigr)  x^i +
\na_{(\alpha}g_{\beta)}(x^{\e})
\,,
\ee
for arbitrary $g_{\beta}(x^{\e})$. Eq. \eqref{sGalilei2} is the string analog of eq.~\eqref{Galilei2}. The action \eqref{nonrelstringpotent} leads to the following modified equations of motion for the transverse coordinates $\{x^i\}$:
\begin{align}\label{geodesic1}
\p_{\bar{\alpha}}& \Bigl( \sqrt{-\bar{\gamma}} \bar{\gamma}^{\bar{\alpha}\bar{\beta}} \p_{\bar{\beta}} x^i \Bigr) + \sqrt{-\bar{\gamma}} \eta^{\alpha\beta} \p^i \Phi_{\alpha\beta}  = 0 \,.
\end{align}
These equations of motion simplify if we choose the static gauge
\be
x^{\alpha}=\sigma^{\bar{\alpha}} \,.\label{stringstaticgauge}
\ee
In this gauge we have that  ${\bar\gamma}_{\bar\alpha\bar\beta} = \eta_{\alpha\beta}$.

The equation of motion of $\Phi_{\alpha\beta}(x)$ itself is easiest obtained by requiring that it is second order in spatial derivatives and invariant
under the acceleration-extended stringy Galilei symmetries \eqref{sGalilei1} and \eqref{sGalilei2}. Since the variation of $\Phi_{\alpha\beta}(x)$,
see eq.~\eqref{sGalilei2}, contains an arbitrary function of the longitudinal coordinates and is linear in the transverse coordinates, it follows that
the unique second-order differential operator satisfying the above requirement is the Laplacian $\Delta \equiv \delta^{ij}\partial_i\partial_j$.
Requiring that the source term is provided by the mass density function $\rho (x)$, which transforms as a scalar with respect to \eqref{sGalilei1}, this leads to the following Poisson equation:
\be\label{Poisson1}
\triangle \Phi_{\alpha\beta} (x) = V_{D-2} G \rho (x) \eta_{\alpha\beta}\,.
\ee
This finishes our first approach where we only gauge the transverse translations. In this approach we have presented both the equations of motion for
the transverse coordinates $\{x^i\}$ of a string, see eq.~(\ref{geodesic1}), as well as the bulk equations of motion for the gravitational potential $\Phi_{\alpha\beta}$, see eq.~(\ref{Poisson1}).

We now proceed with the second gauging procedure in which we gauge the full deformed stringy Galilei algebra. This algebra consists of longitudinal translations, transverse translations, longitudinal Lorentz transformations, ``boost'' transformations, transverse
rotations and two distinct extension transformations.
The explicit commutation relations of the generators corresponding to these symmetries are given in appendix \ref{ExtStringGalalg}.
As a first step one associates a gauge field to each of these symmetries:
\begin{eqnarray}\label{gaugefields}
\tau_\mu{}^a&:&\ \ \ \text{longitudinal translations}\nonumber \\[.1truecm]
e_\mu{}^{a'}&:&\ \ \ \text{transverse  translations}\nonumber \\[.1truecm]
\omega_\mu{}^{ab} &:&\ \ \ \text{longitudinal Lorentz transformations} \\[.1truecm]
\omega_\mu{}^{a'a}&:&\ \ \ \text{``boost'' transformation}\nonumber \\[.1truecm]
\omega_\mu{}^{a'b'}&:&\ \ \ \text{transverse rotations}\nonumber \\[.1truecm]
m_\mu{}^a\,, m_\mu{}^{ab}&:&\ \ \ \text{extension transformations}\,.\nonumber
\end{eqnarray}
 At the same time the constant parameters describing the transformations are promoted to arbitrary functions of the spacetime coordinates $\{x^\mu\}$:
\begin{eqnarray}
\tau^a(x^\mu)&:&\ \ \ \text{longitudinal translations}\nonumber \\[.1truecm]
\zeta^{a^\prime}(x^\mu)&:&\ \ \ \text{transverse translations}\nonumber \\[.1truecm]
\lambda^{ab}(x^\mu) &:&\ \ \ \text{longitudinal Lorentz transformations} \\[.1truecm]
\lambda^{a'a}(x^\mu)&:&\ \ \ \text{``boost'' transformations}\nonumber \\[.1truecm]
\lambda^{a'b'}(x^\mu)&:&\ \ \ \text{transverse rotations}\nonumber \\[.1truecm]
\sigma^a(x^\mu)\,, \sigma^{ab}(x^\mu)&:&\ \ \ \text{extension transformations}\,.\nonumber\label{stringparameters}
\end{eqnarray}
The explicit gauge transformations of the gauge fields, together with the expressions for the
gauge-invariant curvatures and the Bianchi identities they satisfy, can be found in appendix \ref{ExtStringGalalg}. Besides the gauge transformations all gauge fields transform under general coordinate transformations
with parameters $\xi^\mu(x^\mu)= (\xi^\alpha(x^\mu)\,, \xi^i(x^\mu))$.

Like in the particle case we would like to express the $\Gamma$-connection  in terms of the previous gauge fields. In order to do that we first impose a set of so-called conventional constraints on the curvatures of the gauge fields:
\be
R_{\mu\nu}{}^{a}(H) = R_{\mu\nu}{}^{a'}(P) = R_{\mu\nu}{}^{a}(Z) = 0 \,. \label{NHcurvatureconstraints}
\ee
These constraints are required to convert the local $H_a$ and $P_{a^\prime}$ transformations into general coordinate transformations via the identity \eqref{removingHandP}. Besides this, the constraints \eqref{NHcurvatureconstraints} also imply that
the gauge fields $\omega_\mu{}^{a^\prime b^\prime}\,,\ \omega_\mu{}^{a^\prime a}$
and $\omega_\mu{}^{ab}$ become dependent:
\begin{align}
 \omega_{\mu}{}^{a'b'}
   & =  \partial_{[\mu}e_{\nu]}{}^{a'} e^{\nu\,b'} - \partial_{[\mu}e_{\nu]}{}^{b'} e^{\nu\,a'}
         + e_{\mu}{}^{c'} \partial_{[\nu}e_{\rho]}{}^{c'}  e^{\nu\,a'}e^{\rho\,b'}
       - \tau_{\mu}{}^a e^{\rho\,[a'}\omega_{\rho}{}^{b']a} \,,  \label{NHomega2}\\
       &\nonumber \\
\omega_{\mu}{}^{a'a}
& =  2\tau_{\mu}{}^b\Bigl( \tau^{\nu b}e^{\rho a'}[\p_{[\nu}m_{\rho ]}{}^a - \omega_{[\nu}{}^{ac}m_{\rho ]}{}^c] -e^{\nu a'}m_{\nu}{}^{ab} \Bigr)  \nonumber\\
& + 2 e_{\mu}{}^{b'}\tau^{\rho a}e^{\nu (b'}\p_{[\nu}e_{\rho ] }{}^{a')} +
 e_{\mu}{}^{b'} e^{\nu b'} e^{\rho a'}[\p_{[\nu} m_{\rho ]}{}^a -\omega_{[\nu}{}^{ab}m_{\rho ]}{}^b] \,, \label{NHomega3}\\
 &\nonumber \\
\omega_{\mu}{}^{ab} & = \partial_{[\mu}\tau_{\nu]}{}^a\tau^{\nu b} - \partial_{[\mu}\tau_{\nu]}{}^b\tau^{\nu a} + \tau^{\nu a}\tau^{\rho b} \tau_{\mu}{}^c \partial_{[\nu}\tau_{\rho]}{}^c \,.  \label{NHomega1}
\end{align}
The solution for $\omega_{\mu}{}^{ab}$ is familiar from the Poincar\'{e} theory and reflects the fact that the foliation space is given by a two-dimensional Minkowski spacetime. The same constraints have a third effect, namely that they lead to constraints on the curl of the
gauge field $\tau_\mu{}^a$. More precisely, the conventional constraint $R_{\mu\nu}{}^a(H)=0$ can
not only be used to solve for the spin connection $\omega_\mu{}^{ab}$, see eq.~\eqref{NHomega1}. Substituting this solution back into the constraint also implies that the following projections of $\p_{[\mu}\tau_{\nu]}{}^a$ vanish:
\begin{equation}
 e^{\mu a'} \tau^{\nu (a}\p_{[\mu}\tau_{\nu]}{}^{b)}   = 0 \,, \hskip 2truecm
 e^{\mu}{}_{a'} e^{\nu}{}_{b'} \p_{[\mu}\tau_{\nu]}{}^a  = 0 \,. \label{additionalvielbeinconstraints}
\end{equation}

It is instructive to verify how the other two spin connections are solved for. First, the conventional constraints $R_{\mu\nu}{}^{a^\prime}(P)=0$
can not only be used to solve for the spin connection $\omega_\mu{}^{a^\prime b^\prime}$,
see eq.~\eqref{NHomega2}, but also for the
following projections of the spin connection field $\omega_\mu{}^{a^\prime a}$:
\begin{equation}
e^{\mu (a'}\omega_{\mu}{}^{b')b}= 2 \tau^{\nu b} e^{\mu (a'} \p_{[\mu}e_{\nu]}{}^{b')}   \,, \hskip 1.5truecm
\omega_{\rho}{}^{a'[a}\tau^{b]\rho} = -\tau^{\mu a} \tau^{\nu b} \p_{[\mu}e_{\nu]}{}^{a'}   \,. \label{solvedpartsofomegaa'a1}
\end{equation}
Making different contractions of the third conventional constraint $R_{\mu\nu}{}^{a}(Z) = 0$ one
can solve for two more projections of the same spin connection field:
\begin{align}
\tau^{\mu b}\omega_{\mu}{}^{a'a} & = 2 \tau^{\mu b}e^{\nu a'}\Bigl( \p_{[\mu}m_{\nu]}{}^a - \omega_{[\mu}{}^{ac}m_{\nu]}{}^c  \Bigr) - 2e^{\mu a'}m_{\mu}{}^{ab} \,, \label{solvedpartsofomegaa'a2}\\[.15truecm]
e^{\mu [a'}\omega_{\mu}{}^{b']a} & = e^{\mu a'} e^{\nu b'}\Bigl( \p_{[\mu}m_{\nu]}{}^a -\omega_{[\mu}{}^{ab}m_{\nu]}{}^b \Bigr)\,. \label{solvedpartsofomegaa'a3}
\end{align}
Combining the solutions \eqref{solvedpartsofomegaa'a1}, \eqref{solvedpartsofomegaa'a2} and \eqref{solvedpartsofomegaa'a3} for the different projections and using the decomposition
\begin{equation}
\omega_{\mu}{}^{a'a}  = \tau_{\mu}{}^b \tau^{\nu b}\omega_{\nu}{}^{a'a} + e_{\mu}{}^{b'} e^{\nu (b'}\omega_{\nu}{}^{a')a} +
e_{\mu}{}^{b'}e^{\nu [b'}\omega_{\nu}{}^{a']a}\,,
\end{equation}
one can solve for the spin connection field $\omega_\mu{}^{a^\prime a}$, see \eqref{NHomega3}. Finally, it turns out that beyond the
contractions already considered there is one more contraction of the conventional constraint $R_{\mu\nu}{}^{a}(Z) = 0$. It leads to the following constraint on the gauge field $m_\mu{}^{ab}$\,:
\begin{align}
\tau^{\mu [c}m_{\mu}{}^{d]a} & = \tau^{\mu c}\tau^{\nu d} \Bigl( \p_{[\mu}m_{\nu]}{}^a - \omega_{[\mu}{}^{ab}m_{\nu]}{}^b\Bigr) \,.\label{conditionsofR(Z)}
\end{align}
This constraint relates the longitudinal projection of $D_{[\mu}m_{\nu]}{}^a$ to a certain projection of the gauge field $m_{\mu}{}^{ab}$, but does not allow one to solve $m_{\mu}{}^{ab}$ completely; the other projections remain unspecified. We  will return to the meaning of the constraint \eqref{conditionsofR(Z)} after eq.\eqref{NKHshifttautau}.

At this point the symmetries of the theory are the general coordinate transformations, the longitudinal Lorentz
transformations, ``boost'' transformations, transverse rotations and extension transformations,
all with parameters that are  arbitrary functions of the spacetime coordinates. The gauge fields
$\tau_\mu{}^a$ of longitudinal translations and $e_\mu{}^{a^\prime}$ of transverse translations are
identified as the (singular) longitudinal and transverse vielbeins. One may also introduce their inverses (with respect to the longitudinal and transverse subspaces) $\tau^\mu{}_a$ and $e^\mu{}_{a^\prime}$:
\begin{eqnarray}
e_{\mu}{}^{a^\prime} e^{\mu}{}_{b^\prime} & = &\delta_{b^\prime}^{a^\prime}, \ \ \ \ \ e_{\mu}{}^{a^\prime} e^{\nu}{}_{a^\prime}  = \delta^{\nu}_{\mu} - \tau_{\mu}{}^a\tau^{\nu}{}_a\,,\ \ \ \ \ \tau^{\mu}{}_a\tau_{\mu}{}^b = \delta_a^b\,,\nonumber\\[.15truecm]
\tau^{\mu}{}_a e_{\mu}{}^{a^\prime} & = &0, \ \ \ \ \ \ \ \tau_{\mu}{}^a e^{\mu}{}_{a^\prime} =  0\,.
\end{eqnarray}
The spatial and temporal vielbeins are related to the spatial metric $h^{\mu\nu}$ with ``inverse'' $h_{\mu\nu}$, and the temporal metric $\tau_{\mu\nu}$ with ``inverse'' $\tau^{\mu\nu}$, as follows:
\begin{eqnarray}
\tau_{\mu\nu} &=& \tau_\mu{}^a \tau_\nu{}^b\, \eta_{ab}\,, \hskip 1.8truecm \tau^{\mu\nu} = \tau^\mu{}_a\tau^\nu{}_b\,\eta^{ab}\,,\nonumber\\ [.15truecm]
h_{\mu\nu} &=& e_\mu{}^{a^\prime}e_\nu{}^{b^\prime}\,\delta_{a^\prime b^\prime}\,,\hskip 1.5truecm
h^{\mu\nu} = e^\mu{}_{a^\prime}e^\nu{}_{b^\prime}\,\delta^{a^\prime b^\prime}\,. \label{relationNCmetricsvielbeins}
\end{eqnarray}
These tensors satisfy the Newton-Cartan metric conditions
\begin{align}
h^{\mu\nu}h_{\nu\rho} + \tau^{\mu\nu}\tau_{\nu\rho} & = \delta^{\mu}_{\rho}, \ \ \tau^{\mu\nu}\tau_{\mu\nu} = 2 \,, \nonumber\\
h^{\mu\nu}\tau_{\nu\rho} = h_{\mu\nu}\tau^{\nu\rho} & = 0 \,. \label{NC htauconditions}
\end{align}
We note that for the point particle one would have $\tau^{\mu\nu}\tau_{\mu\nu} = 1$ instead of $\tau^{\mu\nu}\tau_{\mu\nu} = 2$.

A $\Gamma$-connection can be introduced by imposing the following  vielbein postulates:
\begin{align}
\partial_{\mu} & e_{\nu}{}^{a'} - \omega_{\mu}{}^{a'b'}e_{\nu}{}^{b'} - \omega_{\mu}{}^{a'a}\tau_{\nu}{}^{a} - \Gamma^{\lambda}_{\nu\mu} e_{\lambda}{}^{a'} = 0 \,, \nonumber\\[0.15truecm]
\partial_{\mu}& \tau_{\nu}{}^a - \omega_{\mu}{}^{ab}\tau_{\nu}{}^b - \Gamma^{\rho}_{\nu\mu} \tau_{\rho}{}^a = 0 \,.
\end{align}
These vielbein postulates allow one to solve for $\Gamma$ uniquely. The torsion $\Gamma^{\rho}_{[\nu\mu]}$ vanishes because of the constraints $R(P)=R(H)=0$, and with this the vielbein postulates give the solution
\begin{align}
\Gamma^{\rho}_{\nu\mu} & = \tau^{\rho}{}_{a} \Bigl(\partial_{(\mu}\tau_{\nu)}{}^a - \omega_{(\mu}{}^{ab}\tau_{\nu)}{}^b \Bigr)	
 + e^{\rho}{}_{a'} \Bigl(	\partial_{(\mu}e_{\nu)}{}^{a'} - \omega_{(\mu}{}^{a'b'}e_{\nu)}{}^{b'}  - \omega_{(\mu}{}^{a'a}\tau_{\nu)}{}^{a}\Bigr)  \label{NHGammastr}
\end{align}
in terms of the dependent spin connections $\omega_\mu{}^{a^\prime b^\prime}\,,\omega_\mu{}^{a^\prime a}$ and $\omega_\mu{}^{ab}$. If one plugs in the explicit solutions of these spin connections, one obtains
\begin{eqnarray}\label{NHcovariantconnection}
\Gamma^{\rho}_{\mu\nu} & = &
\frac{1}{2}\tau^{\rho\sigma} \Bigl(\p_{\nu}\tau_{\sigma\mu} + \p_{\mu}\tau_{\sigma\nu} - \p_{\sigma}\tau_{\mu\nu}\Bigr)
+ \frac{1}{2}h^{\rho\sigma} \Bigl(\p_{\nu}h_{\sigma\mu} + \p_{\mu}h_{\sigma\nu} - \p_{\sigma}h_{\mu\nu}\Bigr)\nonumber  \\ [.1truecm]
&&+\ h^{\rho\sigma}K_{\sigma(\mu}{}^a \,\tau_{\nu)}{}^a,
\end{eqnarray}
where $K_{\mu\nu}{}^a = -K_{\nu\mu}{}^a$ is given by the covariant curl of $m_\mu{}^a$:
\be
K_{\mu\nu}{}^a = 2 D_{[\mu}m_{\nu]}{}^a \,. \label{KisDM}
\ee
An important observation is that $m_\mu{}^{ab}$ does not appear in \eqref{NHcovariantconnection}. The origin of this absence is the fact that the expression \eqref{NHGammastr} is invariant under the shift transformations
\be
\omega_{\mu}{}^{a'a} \rightarrow \omega_{\mu}{}^{a'a} + \tau_{\mu}{}^b X^{a'}_{ab} \,, \label{omegaa'ashift}
\ee
where $X^{a'}_{ab} = X^{a'}_{[ab]}$ is an arbitrary shift parameter. The field $m_\mu{}^{ab}$ appears in the form $X^{a'}_{ab} = e^{\lambda}{}_{a'}m_{\lambda}{}^{ab}$ in the solution of $\omega_\mu{}^{a'a}$, and as such $m_\mu{}^{ab}$ will drop out of the connection \eqref{NHGammastr}, and thus out of \eqref{NHcovariantconnection}.

The Riemann tensor can be obtained, using the vielbein postulates, from the curvatures of the spin connection fields:
\begin{equation}
R^{\mu}_{\ \nu\rho\sigma}(\Gamma)  = - \tau^{\mu}{}_a R_{\rho\sigma}{}^{ab}(M)\tau_{\nu}{}^b - e^{\mu}{}_{a'}R_{\rho\sigma}{}^{a'b'}(M'')e_{\nu b'} - e^{\mu}{}_{a'}R_{\rho\sigma}{}^{a'a}(M') \tau_{\nu a} \,. \label{NHRiemanntensor}
\end{equation}
Note that this Riemann tensor has no dependence on the gauge field $m_\mu{}^{ab}$.

At this stage the independent fields are given by $\{\tau_{\mu}{}^a, e_{\mu}{}^{a'}, m_{\mu}{}^a\}$, whereas we saw that $m_{\mu}{}^{ab}$ was partially solved for via eq.~\eqref{conditionsofR(Z)} and does not enter the
dynamics.\footnote{ An analogous results holds for the dynamics of the non-relativistic string, see eq.~(32) of  \cite{Brugues:2006yd}.} The dynamics of a Newton-Cartan string is now described by the following Lagrangian:
\begin{equation}
L = - \frac{T}{2} \sqrt{-\text{det}(\tau)} \tau^{\bar{\alpha}\bar{\beta}}\p_{\bar{\alpha}} x^{\mu} \p_{\bar{\beta}} x^{\nu} \Bigl( h_{\mu\nu} - 2 m_{\mu}{}^a \tau_{\nu}{}^a \Bigr) \,, \label{altstringaction}
\end{equation}
where the induced  world-sheet metric $\tau_{\bar{\alpha}\bar{\beta}}$ is given by
\be
\tau_{\bar{\alpha}\bar{\beta}} \equiv \p_{\bar{\alpha}} x^{\mu} \p_{\bar{\beta}} x^{\nu} \tau_{\mu\nu} \,. \label{covariantizedwsmetric}
\ee
Eq.~\eqref{altstringaction} is the stringy generalization of the particle action \eqref{alternative}. The first term in eq.~\eqref{altstringaction} can be seen as the covariantization of the Lagrangian of \eqref{nonrelaction} with the Newton-Cartan metrics $h_{\mu\nu}$ and $\tau_{\mu\nu}$, where the induced world-sheet metric \eqref{covariantizedwsmetric} is the covariantization of \eqref{NRpullbacklongmetric} with $\tau_{\mu\nu}$. Analogously to the point particle, the Lagrangian \eqref{altstringaction} is quasi-invariant under the gauged deformed stringy Galilei algebra. Under $Z_a$-transformations $\delta m_{\mu}{}^a=\p_{\mu}\sigma{}^a$ the Lagrangian \eqref{altstringaction} transforms as a total derivative, while  the other transformations leave the Lagrangian invariant. In particular, this applies to the  $Z_{ab}\ $-transformations which
are given by
\be
\delta m_{\mu}{}^a = - \sigma^{ab}\tau_{\mu}{}^b\hskip .5truecm \text{or}\hskip .5truecm \tau^{\mu [a} \delta m_{\mu}{}^{b]} = \sigma^{ab} \,. \label{Zabtransfoofmmua}
\ee
The latter way of writing shows that the
projection $\tau^{\mu [a}  m_{\mu}{}^{b]}$ of the gauge field $m_\mu{}^a$ can be gauged away. The
 $m_{(\mu}{}^a \tau_{\nu)}{}^a$ term  in the Lagrangian \eqref{altstringaction} is needed in order to render the action invariant under boost transformations which transform both the spatial metric $h_{\mu\nu}$ and the extension gauge field $m_\mu{}^a$ as follows:
\be
\delta h_{\mu\nu}=2\lambda^{a'a}e_{(\mu}{}^{a'} \tau_{\nu )}{}^a \,,\hskip 1truecm \delta m_{\mu}{}^a = \lambda^{a'a}e_{\mu}{}^{a'} \,.
\ee
Like in the particle case, the presence of the extension gauge field  $m_\mu{}^a$ represents an ambiguity when trying to solve the $\Gamma$-connection in terms of the (singular) metrics \eqref{relationNCmetricsvielbeins} of Newton-Cartan spacetime. Namely, the metric compatibility conditions on $h^{\mu\nu}$ and $\tau_{\mu\nu}$ give the solution \eqref{NHcovariantconnection}, but $K_{\mu\nu}{}^a = -K_{\nu\mu}{}^a$ is an ambiguity which is not fixed by the metric compatibility conditions. It is the specific solution \eqref{NHGammastr} of the vielbein postulates
which fixes this ambiguity  to be \eqref{KisDM}. A new feature of the string case is that the ambiguity
$K_{\mu\nu}{}^a$ has its own ambiguity. In other words there is an ambiguity in the ambiguity! To show how this works we first note that
from eq.~\eqref{NHcovariantconnection} it follows that the longitudinal projection of \eqref{KisDM} does not contribute to the connection because it is multiplied by $h^{\rho\sigma}$. This is equivalent to saying that the expression  \eqref{NHcovariantconnection} is invariant under the shift
transformations\,\footnote{An analogous result was obtained in \cite{Brugues:2006yd}.}
\be
K_{\mu\nu}{}^a \rightarrow K_{\mu\nu}{}^a + \tau_{[\mu}{}^c \tau_{\nu]}{}^{b} Y^{a}_{bc} \label{NKHshifttautau}
\ee
for arbitrary parameters $Y^{a}_{bc}$. We will now argue that this ambiguity in $K_{\mu\nu}{}^a$ is related to the second extension gauge field, $m_\mu{}^{ab}$, which in contrast to $m_{\mu}{}^a$ does \emph{not} enter the Lagrangian \eqref{altstringaction}. We have
 seen that the absence of $m_\mu{}^{ab}$ in the dynamics follows from  the shift symmetry \eqref{omegaa'ashift}, which prevents the field $m_\mu{}^{ab}$ to enter the $\Gamma$-connection. We now come back to the role of the constraint \eqref{conditionsofR(Z)}
Using eq.~\eqref{KisDM} we see that this constraint relates a certain projection of $m_\mu{}^{ab}$ to the longitudinal projection of the ambiguity $K_{\mu\nu}{}^a$.
This longitudinal projection of the ambiguity is precisely the part that drops out of the expression for $\Gamma$ corresponding to the shift invariance of \eqref{NHcovariantconnection} under \eqref{NKHshifttautau}. Therefore, the constraint \eqref{conditionsofR(Z)} implies that
 a certain projection of the extension gauge field $m_\mu{}^{ab}$ can be regarded as an ``ambiguity in the ambiguity".

 Summarizing, we conclude that the extension gauge field $m_\mu{}^a$, like in the particle case, corresponds to an ambiguity in the $\Gamma$-connection. This
 gauge field occurs in the string action \eqref{altstringaction}. A new feature, absent in the particle case, is that there is a second
 extension gauge field $m_\mu{}^{ab}$ which corresponds to an ambiguity in the ambiguity. This extension gauge field does not occur in the string action \eqref{altstringaction}.

Having clarified the role of the extension gauge fields we now vary the Lagrangian \eqref{altstringaction} which gives, after a long calculation\footnote{Some details are given in appendix \ref{NCgeodesicdetails}.} similar to the one leading to \eqref{geodesiceq},
\be
\tau^{\bar{\alpha}\bar{\beta}} \Bigl(\na_{\bar{\alpha}}\,\p_{\bar{\beta}}\,x^{\rho} + \p_{\bar{\alpha}}x^{\mu} \p_{\bar{\beta}}x^{\nu}\, \Gamma^{\rho}_{\mu\nu} \Bigr) = 0 \,, \label{NCgeodesicstring}
\ee
where the $\Gamma$-connection is given by \eqref{NHGammastr}. This geodesic equation can be seen as the covariantization of \eqref{geodesic1}, and in the particle case reduces to \eqref{geodesiceq} as one would expect. The equations describing the dynamics of stringy Newton-Cartan spacetime are given by
\begin{equation}\label{NCbulk}
R_{\mu\nu}(\Gamma) = V_{D-2}G\rho\tau_{\mu\nu} \,,
\end{equation}
just as for the point particle. The Ricci tensor however now is given in terms of the $\Gamma$-connection \eqref{NHGammastr}.

To make contact with a Galilean observer we impose the additional kinematical constraints
\begin{equation}
R_{\mu\nu}{}^{ab}(M) = R_{\mu\nu}{}^{a^\prime b^\prime}(M^{\prime\prime})=0\,.  \label{constraints2}
\end{equation}
Here $M''$ refers to the generators of spatial rotations, whereas $M$ refers to the generator of a longitudinal rotation which was absent for the particle. It should be stressed that one is not forced to impose these curvature constraints, and one could stay more general and try to solve the resulting theory of gravity for a curved longitudinal and transverse space. In particular, in adding a cosmological constant in the next section, we will impose a different constraint for the longitudinal space. The first constraint of \eqref{constraints2} allows one to gauge-fix $\omega_\mu{}^{ab}=0$, expressing the flatness of the longitudinal space. This solves the
constraints \eqref{additionalvielbeinconstraints} and allows one to go to the so-called adapted coordinates,
in which $\tau_\mu{}^a$ is given by
\begin{equation}\label{adapted2}
\tau_\mu{}^a = \delta_\mu{}^a \,.
\end{equation}
In terms of these adapted coordinates the
(longitudinal and transverse) vielbeins and their inverses are given by
\begin{eqnarray}\label{FlatStringFlatcoord}
\tau_\mu{}^a &=& \bigl(\delta_\alpha^a\,, 0\bigr)\,,\hskip 2truecm e_\mu{}^{a^\prime} = \bigl(- e_k{}^{a^\prime} \tau^k{}_a\,, e_i{}^{a^\prime}\bigr)\,,\nonumber\\[.15truecm]
\tau^\mu{}_a &=& \bigl(\delta^\alpha_a\,, \tau^i{}_{a}\bigr)\,,\hskip 2truecm e^{\mu}{}_{a^\prime}  = \bigl(0\,, e^i{}_{a^\prime}\bigr)\,,
\end{eqnarray}
in terms of the independent components $\tau^i{}_a$ and the transverse  vielbeins $e_i{}^{a^\prime}$ together with their inverse $e^i{}_{a^\prime}$. Note that in adapted coordinates the transverse vielbein is non-singular
in the transverse space, i.e.
\begin{equation}
e_i{}^{a^\prime}\, e^j{}_{a^\prime} = \delta_i^j\,,\hskip 1.5truecm e_i{}^{a^\prime}\, e^i{}_{b^\prime} =
\delta^{a^\prime}_{b^\prime}\,.
 \end{equation}

The second kinematical constraint of \eqref{constraints2} expresses the choice of flat transverse directions. It implies, using
eq.~\eqref{NHRiemanntensor}, that $R^i{}_{jkl}(\Gamma)=0$ and allows us to choose a flat Cartesian coordinate system in the transverse space such that
\begin{equation}
e_i{}^{a^\prime} = \delta_i{}^{a^\prime}\,,\hskip 2truecm e^i{}_{a^\prime} = \delta^i{}_{a^\prime}\,.
\end{equation}
As such the constraints \eqref{constraints2} can be regarded as metric ans\"{a}tze in which one is looking for solutions of the metrics describing both a flat transverse space and a flat foliation space. All metric components can now be expressed in terms of the only nontrivial components $\tau^i{}_a$:
\begin{eqnarray}\label{expressions}
\tau_\mu{}^a &=& \bigl(\delta_\alpha^a\,, 0\bigr)\,,\hskip 2truecm e_\mu{}^{a^\prime} = \bigl(-  \tau^{a^\prime}{}_a\,, \delta_i{}^{a^\prime}\bigr)\,,\nonumber\\[.15truecm]
\tau^\mu{}_a &=& \bigl(\delta^\alpha_a\,, \tau^i{}_{a}\bigr)\,,\hskip 2truecm e^{\mu}{}_{a^\prime}  = \bigl(0\,, \delta^i{}_{a^\prime}\bigr)\,,
\end{eqnarray}
where
we do not distinguish anymore between (longitudinal, transverse) curved indices $(\alpha,i)$ and (longitudinal, transverse) flat indices $(a,a^\prime)$.

Plugging the conventional constraints \eqref{NHcurvatureconstraints} and the kinematical constraints \eqref{constraints2}
into the Bianchi identities \eqref{Bianchi} we find that
\begin{equation}
R_{\alpha\beta}(\Gamma) = -\delta_{(\alpha}^a\delta_{\beta)}^b e^\rho{}_{a^\prime}\tau^\sigma{}_b R_{\rho\sigma}{}^{a^\prime a}(M^\prime) \label{NHnonzeroRicci}
\end{equation}
are the only nonzero components of the Ricci tensor. Furthermore, the remaining nonzero curvatures $R(M^\prime)$ and $R(Z)$ are constrained by the following algebraic identities:
\be
R_{[\lambda\mu}{}^{a'a}(M')\tau_{\nu]}{}^a  = R_{[\lambda\mu}{}^{a'a}(M')e_{\nu]}{}^{a'}
-R_{[\lambda \mu}{}^{ab}(Z)\tau_{\nu]}{}^b = 0 \,. \label{NHremainingBianchis}
\ee

The kinematical constraint $R_{\mu\nu}{}^{a^\prime b^\prime}(M^{\prime\prime})=0$  also allows one to gauge-fix $\omega_\mu{}^{a^\prime b^\prime}=0$. We will now show that in this gauge
\begin{equation}
\Gamma^i_{\alpha j}=0\,,\hskip 1.5truecm \Gamma^i_{\alpha\beta} = \partial^i\Phi_{\alpha\beta}\,,
\end{equation}
where the latter equation defines the gravitational potential $\Phi_{\alpha\beta}$.

We first show that $\Gamma^i_{\alpha j}=0$.
Using the expressions \eqref{expressions}, eq.~\eqref{NHGammastr} and the fact that $\omega_j{}^{ab} = \omega_\mu {}^{a^\prime b^\prime}=0$ we find that  $\Gamma^i_{\alpha j}$
is given by
\begin{equation}
\Gamma^i_{a j} = \frac{1}{2}\bigl(-\partial_j\tau^i{}_a -\omega_j{}^{ia} \bigr)\,.
\end{equation}
Next, using expressions \eqref{NHomega2}-\eqref{NHomega1}, we find that
\begin{eqnarray}
\omega_j{}^{ia}  = -\partial_{[i}m_{j] a} - \partial^{(i}\tau^{j)a}\,,
\end{eqnarray}
where we have used that $\omega_i{}^{ab}=0$. Furthermore, the gauge-fixing condition  $\omega_k{}^{ij}=0$ is already satisfied but the
gauge-fixing condition $\omega_\alpha{}^{a^\prime b^\prime}=0$ leads to the constraint
\begin{eqnarray}
\omega_a{}^{ij} = -\partial_{[i}m_{j] a} - \partial^{[i}\tau^{j]a} =0\,.\label{constraint11}
\end{eqnarray}
This constraint equation implies that $m_{ia}$ can be
written as
\begin{equation}\label{total}
m_{ia} = - \tau^{i}{}_{a} -\partial_i m_a\,,
\end{equation}
where $m_a$ are the transverse spatial gradient components of $m_{ia}$.
Substituting the expression for $\omega_j{}^{ia}$ into that of $\Gamma^i_{aj}$ the result becomes proportional to the
righthand-side of the constraint equation \eqref{constraint11} and hence we find $\Gamma^i_{aj}=0$.

We next show that $\Gamma^i_{\alpha\beta}$ can be written as $\partial^i\Phi_{\alpha\beta}$  defining a gravitational potential $\Phi_{\alpha\beta}$.
Using \eqref{NHGammastr} we derive the following expression:\,\footnote{Remember that we do not distinguish anymore between flat indices $a$ and curved indices $\alpha$.}
\begin{equation}
\Gamma^i_{ab} = -\partial_{(a}\tau^i{}_{b)} - \omega_{(a}{}^i{}_{b)}\,, \label{StringGalileiGammaiab}
\end{equation}
where we have used that $\omega_\alpha{}^{ab} = \omega_\alpha{}^{ij}=0$.
Following eqs.~\eqref{NHomega2}-\eqref{NHomega1} we find that $\omega_a{}^{ib}$ is given by
\begin{equation}
\omega_a{}^{ib} = \partial_a m_{ib}-\partial_i m_{ab} + \tau^k{}_a\partial_{[k}m_{i]b} +\frac{1}{2}\tau^k{}_a\bigl(\partial_i\tau^k{}_b\bigr)
+\frac{1}{2}\tau^k{}_a\partial_k\tau^i{}_b +2 m_i{}^{ab}\,.
\end{equation}
Substituting this expression for $\omega_a{}^{ib}$ back into that of $\Gamma^i_{ab}$ and using \eqref{total} we indeed find that $\Gamma^i_{ab} = \partial^i\Phi_{ab}$ with
\begin{equation}
\Phi_{\alpha\beta}(x) = m_{(\alpha\beta)}(x) -\frac{1}{2}\delta_{ij}\tau^i{}_\alpha(x)\tau^j{}_\beta (x) +\partial_{(\alpha}m_{\beta)}(x)\,, \label{phiabstringncalgebraic}
\end{equation}
where $m_{(\alpha\beta)} = m_{(\alpha}{}^a \delta^a_{\beta)}$. This is the stringy generalisation of eq.~\eqref{gravpot}.

Using the expressions for the components of the $\Gamma$-connection calculated above we may now verify  that the Newton-Cartan geodesic equation \eqref{NCgeodesicstring} and the Poisson equation \eqref{NCbulk} corresponding to the second gauging procedure reduce to the equations  \eqref{geodesic1} and \eqref{Poisson1}
derived in the first gauging procedure. After gauge-fixing the Newton-Cartan symmetries to the accele\-ration-extended Galilei symmetries as described above,  the Lagrangian \eqref{altstringaction} reduces to the Lagrangian associated to the action \eqref{nonrelstringpotent}, with the potential $\Phi_{\alpha\beta}$ given by \eqref{phiabstringncalgebraic} and $\bar{\gamma}_{\bar{\alpha}\bar{\beta}} = \tau_{\bar{\alpha}\bar{\beta}}$:\footnote{After the gauge-fixing one has $\tau_{\bar{\alpha}\bar{\beta}} = \p_{\bar{\alpha}} x^{\alpha} \p_{\bar{\beta}}x^{\beta}\eta_{\alpha\beta}$.}
\be
L = - \frac{T}{2} \sqrt{-\text{det}(\tau)} \, \tau^{\bar{\alpha}\bar{\beta}} \Bigl( \p_{\bar{\alpha}} x^i \p_{\bar{\beta}} x^j \delta_{ij} + \p_{\bar{\alpha}} x^{\alpha} \p_{\bar{\beta}} x^{\beta}[\tau^i{}_{\alpha} \tau^{j}{}_{\beta}\delta_{ij} - 2 m_{(\alpha\beta)} - 2 \p_{(\alpha}m_{\beta)}] \Bigr)\,.
\ee
The longitudinal components $R_{\alpha\beta}(\Gamma)$ of the Ricci tensor become
\be
R_{\alpha\beta}(\Gamma) = -\delta_{(\alpha}^a \delta_{\beta)}^b e^\rho{}_{a^\prime}\tau^\sigma{}_b R_{\rho\sigma}{}^{a^\prime a}(M^\prime) = \delta^{ij}\p_i\p_j \Phi_{\alpha\beta} \,,
\ee
such that indeed \eqref{NCbulk} gives the stringy Poisson equation \eqref{Poisson1}. This finishes our discussion of the string moving in a flat Minkowski spacetime. In the next section we will consider the addition of a cosmological constant.

\section{Adding a Cosmological Constant}

In order to study  applications of the AdS/CFT correspondence based on the symmetry algebra corresponding to a non-relativistic string
it is necessary to include a (negative) cosmological constant $\Lambda$. To explain how this can be done,
we will discuss in the first subsection the particle case. In the second subsection we
will show how to go from particles to strings.

\subsection{The Particle Case}

Adding a negative cosmological constant in the relativistic case means that  the Poincar\'e algebra gets replaced by an Anti-de Sitter (AdS) algebra corresponding to a particle
moving in an $\text{AdS}$  background. It is well-known that  one cannot obtain general relativity with a (negative) cosmological constant by gauging the AdS algebra in the same way that one can obtain general relativity by gauging the Poincar\'e algebra \cite{book}. The (technical) reason for this is that one cannot
find a set of (so-called conventional) curvature constraints whose effect is to convert the translation transformations into
general coordinate transformations and, at the same time, to make certain gauge fields to be dependent on others. The same is true for the non-relativistic limit of the AdS algebra which is the
Newton-Hooke algebra \cite{Bacry:1968zf,Gibbons:2003rv}.
Therefore, we cannot apply the same gauging procedure to the Newton-Hooke algebra that we used for the  Bargmann algebra in section 2.
It turns out that we do not need to apply a full gauging procedure to the Newton-Hooke algebra.
When taking the non-relativistic limit of a particle moving in an AdS background, which is a $\Lambda$-deformation
of the Minkowski background, one ends up with the action of a non-relativitic particle moving in a harmonic oscillator potential.
This is a particular case of the non-relativistic
particle action for a Galilean observer with {\sl zero} cosmological constant but with a
particular  non-zero-value of the
potential $\Phi(x)$. In view of this it is convenient to write the potential $\Phi(x)$ as the sum of a
purely gravitational potential $\phi(x)$ and an effective background potential $\phi_\Lambda(x)$ describing the harmonic oscillator
due to the cosmological constant:
\begin{equation}\label{split}
\Phi(x) =  \phi(x) + \phi_\Lambda(x)\,.
\end{equation}
Notice that eq.\eqref{split} points out a conceptual difference between the relativistic and non-relativistic notion of a cosmological constant, which will also be true for the string. Namely, according to \eqref{split} one is always able to redefine the potential $\phi(x)$ in order to absorb the cosmological constant into $\Phi(x)$. But in the relativistic case such a redefinition of $\Lambda$ into the metric $g_{\mu\nu}(x)$ is not possible.
The non-relativistic particle action in the presence of a cosmological constant is invariant under the Newton-Hooke symmetries which is a $\Lambda$-deformation of the Galilei symmetries we considered in section 2. A particularly useful feature of the Newton-Hooke symmetries is that the
$\Lambda$-deformed symmetries can all be viewed
as particular time-dependent transverse translations. This means that, when gauging the Galilei symmetries
like we did in section 2, the Newton-Hooke symmetries are automatically included. The consequence of this is that, although we cannot perform the second gauging procedure of section 2, i.e.~gauge the full
Newton-Hooke algebra, it is straightforward to apply the first gauging procedure, i.e.~gauge the transverse
translation leading to  arbitrary accelerations between different frames, as is appropriate
for a Galilean observer. Independent of whether we are starting from the Galilei or Newton-Hooke
symmetries, when we gauge the transverse translations we end up with precisely the same answer
which we already derived in section 2, but with a different interpretation of the potential $\Phi(x)$.
The difference is seen when we turn off gravity. Without a cosmological constant, turning off
gravity means setting $\Phi(x)=\phi(x)=0$ and there is no background potential, i.e.~$\phi_\Lambda(x)=0$.
However, when $\Lambda\ne 0$, turning off gravity means a different thing since now we want to end up
with a non-zero background potential $\phi_\Lambda(x)\ne 0$. According to eq.~\eqref{split} it means
setting $\Phi(x) = \phi_\Lambda(x)$ or $\phi(x)=0$. One can view this as a different gauge condition
and that is the reason why, in the presence of a non-zero cosmological constant, the symmetries that
 relate inertial frames is given by the Newton-Hooke
symmetries instead of the Galilei symmetries. For a Galilean observer, however, we end up with precisely the same geodesic equation and bulk equation of motion we derived in the absence of a cosmological
constant in the previous section.

Before showing how the Newton-Hooke symmetries arise as the transformations that relate inertial frames, it is instructive to first re-derive the Galilei symmetries starting from a Galilean observer. Consider  the acceleration-extended Galilei symmetries given in
eqs.~\eqref{Galilei1} and \eqref{Galilei2}. Without a cosmological constant, turning off gravity means
setting $\Phi(x)=0$. Given the transformation rule \eqref{Galilei2} of the background potential $\Phi(x)$
this implies the following restriction on the transverse translations:
\begin{equation}\label{restriction1}
\frac{d}{d\tau} \Bigl(\frac{\dot{\xi}^i}{\dot{t}}\Bigr)=0 \,,
\end{equation}
where we have ignored the standard ambiguity in the potential represented by the function $g(t)$
in eq.~\eqref{Galilei2}.
This restriction implies that ${\dot\xi^i} = v^i {\dot t}$ or $\xi^i(t) = v^i t +  \zeta^i$.
This brings us back to the
Galilei transformations given in eq~\eqref{Galileisymmetries}.

We now turn to the case of a non-zero cosmological constant $\Lambda$. It turns out that, when taking the non-relativistic limit as is described in section \ref{The Particle Case} of a particle moving in an (A)dS background,\footnote{For this the cosmological constant $\Lambda$ must be rescaled with a factor of $\omega^{-2}$.} one ends up with a particle moving in an effective background potential $\phi_\Lambda = -\tfrac{1}{2}\Lambda x^i x^i$ describing a harmonic oscillator \cite{Bacry:1968zf}:
\begin{equation}
S = \frac{m}{2}\int \Bigl( \frac{\dot{x}^i \dot{x}^j\delta_{ij}}{\dot{t}} + \dot{t}\, \Lambda x^i x^j \delta_{ij} \Bigr)d\tau \,. \label{NHparticle}
\end{equation}
We take the convention in which $\Lambda > 0$ describes a dS space, whereas $\Lambda < 0$ gives an AdS space. In the following we will consider  the AdS case only. The action \eqref{NHparticle} is nothing else than the action \eqref{actionparticle}, with $\Phi(x)$ being the harmonic oscillator potential,
\be
\Phi(x) = \phi_\Lambda(x) = -\tfrac{1}{2}\Lambda x^i x^i  \,. \label{NHparticleBG}
\ee
Viewed as a gauge condition, and using the transformation rule \eqref{Galilei2}, this equation  is invariant under transverse translations that satisfy the following constraint:
\begin{equation}\label{restriction2}
\frac{1}{\dot t}\frac{d}{d\tau} \Bigl(\frac{\dot{\xi}^i}{\dot{t}}\Bigr)=
\Lambda \xi^i \,.
\end{equation}
Here we have again ignored the ambiguity in the potential represented by the function $g(t)$
in eq.~\eqref{Galilei2}.
For $\Lambda<0$, i.e. AdS space,  the restriction \eqref{restriction2} on $\xi^i$ is solved by\footnote{For $\Lambda > 0$, i.e. dS space, one obtains a similar expression but with  the sine and cosine
replaced by their hyperbolic counterparts.}
\begin{equation}
\xi^i (t) = v^i R \sin{(\frac{t}{R})} + \zeta^i \cos{(\frac{t}{R})} \,, \label{NHxi}
\end{equation}
where
\be
R^2 \equiv -\frac{1}{\Lambda} \,.
\ee
Note that for $\Lambda \rightarrow 0$ or $R\rightarrow \infty$ this expression reduces to the Galilei result
$\xi^i(t) = v^i t +  \zeta^i$.

The complete transformation rules are now obtained by combining the transformations
\eqref{NHxi} with the constant time translations and the spatial rotations:
\be\label{NHpointparticlesymmetries}
\delta t = \zeta^0, \ \ \ \ \delta x^i = \lambda^i{}_j x^j +  v^i R \sin{(\frac{t}{R})} + \zeta^i \cos{(\frac{t}{R})}\,.
\ee
This defines the Newton-Hooke algebra
whose non-zero commutators are given by \cite{Bacry:1968zf} (see also \cite{Gibbons:2003rv}):
\begin{eqnarray}\label{NHalgebra}
[P_{a^\prime},H] &=&  R^{-2}G_{a^\prime}\,,\hskip 1.5truecm
[G_{a^\prime},H]=  - P_{a^\prime}\,,\nonumber\\ [.2truecm]
[M_{a^\prime b^\prime},P_{c^\prime}] &=& -2\eta_{c^\prime[a^\prime}P_{b^\prime]}\,,\hskip 1truecm
[M_{a^\prime b^\prime},G_{c^\prime}] =-2\eta_{c^{\prime}[a^\prime}G_{b^\prime]}\,,\\[.2truecm]
[M_{a^\prime b^\prime},M_{c^\prime d^\prime}] &=& 4\eta_{[a^\prime[c^\prime}M_{d^\prime]b^\prime]}\,.\nonumber
\end{eqnarray}
Here $H, P_{a^\prime},G_{a^\prime}$ and $M_{a^\prime b^\prime}$ are the generators of time translations, spatial translations, boosts and
spatial rotations, with parameters $\zeta^0,\zeta^{a^\prime},v^{a^\prime}$ and $\lambda^{a^\prime b^\prime}$, respectively.
We note that the cosmological constant shows up in the $[P_{a^\prime},H]$ commutator, but not in the $[P_{a'},P_{b'}]$ commutator. This is consistent with  the fact that the transverse space is flat. We also
 observe that at this stage the Newton-Hooke algebra \eqref{NHalgebra} does not contain a central extension like the Bargmann algebra,
 i.e.~$[P_{a^\prime},G_{b^\prime}]=0$. Similar to the Galilei particle action \eqref{nrparticlereparaminvaction} the Newton-Hooke particle action \eqref{NHparticle} suggests a central extension: the corresponding Lagrangian is quasi-invariant under both boosts \emph{and} translations, described by the parameter \eqref{NHxi}:
\begin{align}
\delta L & = \frac{d}{d\tau}\Bigl(  \frac{m\, \delta_{ij} x^i\dot{\xi^j}}{\dot{t}}  \Bigr) \nonumber\\
& = \frac{d}{d\tau}\Bigl(mx^i v^j\,\delta_{ij} \cos{(\frac{t}{R})} - mx^i \zeta^j \,\delta_{ij} \sin{(\frac{t}{R})}  \Bigr) \,. \label{NHparticletotalderiv}
\end{align}
This is most easily seen by using the restriction \eqref{restriction2} directly in the variation of the Lagrangian
 corresponding to the action \eqref{NHparticle}. In the limit $R \rightarrow \infty$, i.e.~$\Lambda \rightarrow 0$
 the variation \eqref{NHparticletotalderiv} reduces to the variation \eqref{particleboostquasiinvariance}. Calculating the Noether charges $Q_P$ and $Q_G$ for the translations and the boosts respectively, the Poisson brackets suggest the same central extension $M$ as for the Galilei particle:
\be
[P_{a'},G_{b'}] = \delta_{a'b'} M \,.
\ee
Given the transformation rules \eqref{NHpointparticlesymmetries}, it is straightforward to calculate the commutators between the
different transformations and to verify that they are indeed given by the Newton-Hooke algebra
\eqref{NHalgebra}. As explained above, when viewed as the symmetries of the Newton-Hooke particle described by the action \eqref{NHparticle}, one obtains a centrally-extended Newton-Hooke algebra. The contraction $R \rightarrow \infty$ on this algebra reproduces the Bargmann algebra. This the non-relativistic analog of the fact that the $R \rightarrow \infty$ contraction on the $(A)dS$ algebra yields the Poincar\'{e} algebra.\\

To obtain the cosmological constant in the gauging procedure of the Bargmann algebra we relate the expression for the potential \eqref{gravpot} in terms of the gauge field components to the potential \eqref{split}:
\begin{align}
\Phi (x) & = m_0(x) - \tfrac{1}{2}\delta_{ij}\tau^i (x)\tau^j(x) + \p_0 m(x) \nonumber\\
 & = \phi(x) - \tfrac{1}{2}\Lambda x^i x^j \delta_{ij}   \,.
\end{align}
The Poisson equation \eqref{first2} can then be written as
\be
\triangle \phi (x) = V_{D-2} G \rho (x) + (D-1)\Lambda \,, \label{first2Lambda}
\ee
where $D$ is the dimension of spacetime.


\subsection{The String Case}

We now wish to discuss the string case following the same philosophy as we used for the
particle case in the previous subsection.

Like in the particle case, we write the potential $\Phi_{\alpha\beta}(x)$  as the sum of a purely gravitational potential and a background potential that represents the extra gravitational force represented by the non-zero cosmological constant $\Lambda$:
\begin{equation}
\Phi_{\alpha\beta}(x) = \phi_{\alpha\beta}(x) + \phi_{\alpha\beta,\Lambda}(x)\,. \label{stringpotentialsplit}
\end{equation}
We first consider the case of a zero cosmological constant and show how the stringy Galilei symmetries
are recovered after turning off gravity. According to eq.~\eqref{sGalilei2} the condition $\Phi_{\alpha\beta}(x)=0$ leads
to the following restriction on the transverse translations:
\begin{equation}\label{restrictionL0}
  \p_{\bar{\alpha}}\Bigl( \sqrt{-\bar{\gamma}} \, \bar{\gamma}^{\bar{\alpha}\bar{\beta}}\,\p_{\bar{\beta}} \xi^i  \Bigr)  =0\,,
\end{equation}
where we have ignored the standard ambiguity in $\Phi_{\alpha\beta}(x)$ represented by the
arbitrary functions $g_{\beta}(x^{\e})$ in eq.~\eqref{sGalilei2}.
This restriction  is the stringy analogue
of the restriction \eqref{restriction1} we found in the particle case. It is precisely the same
restriction one finds if one requires that the non-relativistic string action \eqref{nonrelaction} is invariant under transverse translations. The solution of
eq.~\eqref{restrictionL0} is given by $\xi^i(x^\alpha) = \lambda^i{}_\beta x^\beta + \zeta^i$, which can be checked using expression
\eqref{NRpullbacklongmetricinverse}
of $\bar{\gamma}^{\bar{\alpha}\bar{\beta}}$. This brings us back to the stringy Galilei symmetries
given in eq.~\eqref{stringyG}.

We now consider a non-zero cosmological constant $\Lambda$. It turns out that when one considers
the non-relativistic limit of a string moving in an AdS background one ends up with an effective
background potential given by \cite{Gomis:2004pw}
\begin{equation}
\phi_{\alpha\beta,\Lambda} = \tfrac{1}{4}\Lambda x^i x^j \delta_{ij}\tau_{\alpha\beta}\,, \label{stringlambdapotential}
\end{equation}
where $\tau_{\alpha\beta}$ is an ${\rm AdS}_2$-metric. At the same time one should replace the  flat
foliation of spacetime by an ${\rm AdS}_2$-foliation. This means that both in the definition of
${\bar\gamma}_{\bar\alpha\bar\beta}$ given in eq.~\eqref{NRpullbacklongmetric} and the action \eqref{nonrelstringpotent} one should replace the flat metric $\eta_{\alpha\beta}$ by the ${\rm AdS}_2$-metric $\tau_{\alpha\beta}$.  Setting also $\Phi_{\alpha\beta}(x) = \tfrac{1}{4}\Lambda x^i x^j \delta_{ij}\tau_{\alpha\beta}$ in eq.\eqref{nonrelstringpotent}, one obtains the action \cite{Gomis:2004pw}
\begin{align}
S = - \frac{T}{2} \int  d^2 \sigma \sqrt{-\bar{\gamma}} \, \Bigl(\bar{\gamma}^{\bar{\alpha}\bar{\beta}}\p_{\bar{\alpha}} x^i \p_{\bar{\beta}} x^j \delta_{ij} + \Lambda x^i x^j \delta_{ij}  \Bigr) \,,  \label{NHstringcosmoconstant}
\end{align}
with ${\bar\gamma}_{\bar\alpha\bar\beta}$ given by
\begin{equation}
\ \ \  \bar{\gamma}_{\bar{\alpha}\bar{\beta}} = \p_{\bar{\alpha}} x^{\alpha}\p_{\bar{\beta}} x^{\beta} \tau_{\alpha\beta} \,.
\end{equation}
The replacement of $\eta_{\alpha\beta}$ by $\tau_{\alpha\beta}$ also applies to the transformation rule \eqref{sGalilei2}. This leads to the following modified restriction on the transverse translations:
\begin{equation}\label{restrL}
 \frac{1}{\sqrt{-\bar{\gamma}}} \,  \p_{\bar{\alpha}}\Bigl( \sqrt{-\bar{\gamma}} \, \bar{\gamma}^{\bar{\alpha}\bar{\beta}}\,\p_{\bar{\beta}} \xi^i  \Bigr) = -\Lambda \xi^i\,.
 \end{equation}
Note that we have again ignored  the
arbitrary functions $g_{\beta}(x^{\e})$ in eq.~\eqref{sGalilei2}.
For $\Lambda<0$, i.e.~AdS space, the restriction \eqref{restrL} is solved for by the
following expression for $\xi^i(x^\alpha)$\,:
\begin{equation}\label{solL}
\xi^i(x^\alpha) =  \lambda^i{}_0 \sqrt{z^2 + R^2}\sin{(\frac{t}{R})} + \lambda^i{}_1 z +
\zeta^i \frac{\sqrt{z^2 + R^2}}{R} \cos{(\frac{t}{R})}  \,,
\end{equation}
where we have written $x^\alpha = \{t,z\}$ and used that $\Lambda = -R^{-2}$. Note that for $R\rightarrow\infty$ this expression reduces to the stringy Galilei one given by
$\xi^i(x^\alpha) = \lambda^i{}_\beta x^\beta + \zeta^i$.

The complete transformation rules are obtained by combining the transformation rules \eqref{solL}
with the spatial transverse rotations and the isometries of the ${\rm AdS}_2$-space that act on $x^\alpha=\{t,z\}$. The  form of the latter transformations in an explicit coordinate frame is given in appendix \ref{ads2appendix}, see eq.~\eqref{inftz}, where a few useful properties of the ${\rm AdS}_2$ foliation space have been collected. All these transformations together define the stringy Newton-Hooke algebra:
\begin{eqnarray}\label{stringNHalgebra}
[H_a, H_b] & = & R^{-2}M_{ab}\,, \hskip 2truecm [M_{bc} , H_a]  = -2\eta_{a[b}H_{c]}\,,  \nonumber\\ [.2truecm]
[M_{cd} , M_{ef}]  & =& 4 \eta_{[c [e}M_{f]d]}\,, \nonumber\\ [.5truecm]
[P_{a^\prime},H_a] & =& R^{-2}M_{a^\prime a}\,, \hskip 1.5truecm
[M_{c'd'},M_{e'f'}]   = 4 \eta_{[c' [e'}M_{f']d']}\,,  \\ [.2truecm]
[M_{b'c} , H_a]  & = &\eta_{ac}P_{b'}\,, \hskip 2truecm  [M_{b'c'}, P_{a'}]  = -2\eta_{a'[b'}P_{c']}\,,  \nonumber\\ [.2truecm]
[M_{c'd}, M_{ef}] & = & 2 \eta_{d[e}M_{|c'|f]}\,,\hskip 1truecm    [M_{c'd'}, M_{e'f}]  = -2 \eta_{e'[c'}M_{d']f}\,.  \nonumber
\end{eqnarray}
Note that the generators $\{H_a,M_{ab}\}$ span an $\mathfrak{so}(2,1)$ algebra describing the isometries of the ${\rm AdS}_2$-foliation. Using the transformation rules given above and in appendix \ref{ads2appendix} one may calculate the different commutators and verify
that the algebra defined by \eqref{stringNHalgebra} is satisfied. Notice how the cosmological constant ends up in the $[H_a, H_b]$ and $[P_{a^\prime},H_a]$ commutators, but not in the $[P_{a'},P_{b'}]$ commutator. This is consistent with the fact that the transverse space is flat but that the two-dimensional longitudinal space is not flat. Like in the case of the point particle, the stringy Newton-Hooke algebra \eqref{stringNHalgebra} allows for an extension \cite{Gomis:2004pw}. This is motivated by the fact that the Lagrangian $L$ corresponding to the string action \eqref{NHstringcosmoconstant} with the potential \eqref{stringlambdapotential} transforms as a total derivative under the  boosts and translations described by the parameters \eqref{solL}:
\be
\delta L = \p_{\bar{\alpha}} \Bigl(-T \sqrt{-\bar{\gamma}} \,\bar{\gamma}^{\bar{\alpha}\bar{\beta}} x^i \p_{\bar{\beta}} \xi_i \Bigr) \,. \label{NHstringLagrangvar}
\ee
This is most easily seen by using the restriction \eqref{restrL} directly in the variation of the Lagrangian corresponding to  \eqref{NHstringcosmoconstant}.  For $R \rightarrow \infty$ the variation \eqref{NHstringLagrangvar} reduces to the
 variation \eqref{Galstringtotderivative}, and in the particle case it reduces to the variation \eqref{NHparticletotalderiv}. The resulting extension suggested by the Poisson brackets is given by eq.~\eqref{NHdeformation}. \\

We now fit the cosmological constant into the gauging procedure for the string. One important difference with the point particle case is that the foliation space for the string becomes ${\rm AdS}_2$, whereas for the particle this foliation space is trivially flat. To accomplish this ${\rm AdS}_2$-foliation we change the on-shell curvature constraint \eqref{constraints2} for the foliation space, whereas for the transverse space we keep it unaltered:
\be
R_{\mu\nu}{}^{ab}(M) = \Lambda \tau_{[\mu}{}^a \tau_{\nu]}{}^b, \ \ \ R_{\mu\nu}{}^{a'b'}(M'')=0 \,. \label{RMAdS2RM''constraints}
\ee
This gives an ${\rm AdS}_2$ space in the longitudinal direction and a flat transverse space. We then  choose coordinates such that
\begin{align}
\tau_\mu{}^a & = \bigl( \tau_{\alpha}{}^a \,, 0\bigr), \ \ \ \ e_\mu{}^{a^\prime} = \bigl(- \tau^{a'}{}_a \tau_{\alpha}{}^a \,, \delta_i^{a'}\bigr)\,,\nonumber\\[.15truecm]
\tau^\mu{}_a &= \bigl(\tau^\alpha{}_a \,, \tau^i{}_a \bigr), \ \ \ \  e^{\mu}{}_{a^\prime}  = \bigl(0\,, \delta^i_{a^\prime}\bigr)\,, \label{NHAdS2stringcoordinates}
\end{align}
where now we are not able to choose $\tau_{\alpha}{}^a = \delta_{\alpha}^a$, as we did in \eqref{FlatStringFlatcoord}. Using the coordinates chosen in appendix \ref{ads2appendix} one can choose
\begin{eqnarray}
\tau_{\alpha}{}^a &=& \bigl( (1+\frac{z^2}{R^2})^{1/2}\delta_0^a, (1+\frac{z^2}{R^2})^{-1/2} \delta_1^a \Bigr)\,, \\ [.2truecm]
\tau^\alpha{}_a &=& \bigl((1+\frac{z^2}{R^2})^{-1/2}\delta^0_a, (1+\frac{z^2}{R^2})^{1/2} \delta^1_a\Bigr)\,.
 \end{eqnarray}
In view of this we should  carefully  distinguish between the curved longitudinal coordinates $\{\alpha\}$ and the flat longitudinal coordinates $\{a\}$. In contrast, from now on we will not  distinguish between flat and curved transverse coordinates $\{a'\}$ and $\{i\}$ because the transverse space is flat. With the coordinates \eqref{NHAdS2stringcoordinates} the constraints \eqref{RMAdS2RM''constraints} allow for the gauge choice
\be
\omega_{\mu}{}^{a'b'} = 0, \ \ \ \ \omega_{i}{}^{ab} = 0 \,. \label{longandtransvomegaiszero}
\ee
The condition $\omega_{i}{}^{a'b'} = 0$ is trivially satisfied, but an explicit calculation reveals that
\be
\omega_{\alpha}{}^{ij} = -\tau_{\alpha}{}^a \Bigl(\p^{[i}\tau^{j]}{}_a + \p_{[i}m_{j]}{}^a \Bigr) = - \frac{1}{2}\Gamma^{i}_{\alpha j} = 0 \,, \label{omegaalphaijzero}
\ee
so the gauge condition $\omega_{\alpha}{}^{ij}=0$ sets the connection component $\Gamma^{i}_{\alpha j}$ to zero, as in the Galilei string case. From \eqref{omegaalphaijzero} we again arrive at \eqref{total}. One should now be careful in distinguishing between $\tau^i{}_a$, which is nonzero in general, and $\tau_i{}^a$, which is zero for the coordinate choice \eqref{NHAdS2stringcoordinates}. With the spin connections \eqref{longandtransvomegaiszero} and \eqref{omegaalphaijzero} one can show that the expression for the connection, eq.\eqref{NHGammastr}, implies that again $\Gamma^i_{\alpha\beta} = \p^i \Phi_{\alpha\beta}$,
i.e.~the $\Gamma$-connection can also for the ${\rm AdS}_2$-foliation be written as the transverse gradient of a potential. The potential $\Phi_{\alpha\beta}$ is now given by
\be
\Phi_{\alpha\beta} = m_a \omega_{(\alpha}{}^{ab}\tau_{\beta)}{}^b + \tau_{(\alpha}{}^a \p_{\beta)}m_a + \tau_{(\alpha}{}^a m_{\beta)}{}^a - \frac{1}{2}	\tau_{(\alpha}{}^a \tau_{\beta)}{}^b \tau^j{}_a \tau^j{}_b \,, \label{NHphialphabeta}
\ee
which should be compared to the potential for the flat foliation, eq.~\eqref{phiabstringncalgebraic}. To describe the splitting described in the beginning of this section with the background given by \eqref{stringlambdapotential}, we put the potential \eqref{NHphialphabeta} equal to \eqref{stringpotentialsplit}. That the set of gauge fields appearing on the right hand side of \eqref{NHphialphabeta} can give rise to an arbitrary symmetric $\Phi_{\alpha\beta}$ can be seen by taking, for example, the realization $m_a = \tau^i{}_a = 0$ (and thus, via \eqref{total}, $m_i{}^a=0$) in the potential \eqref{NHphialphabeta} and expressing the remaining longitudinal components $m_{\alpha}{}^a$ in terms of $\Phi_{\alpha\beta}$. The symmetric longitudinal projection of $m_{\mu}{}^a$ is then given by
\be
\tau^{\alpha(a} m_{\alpha}{}^{b)} = \tau^{\alpha a} \tau^{\beta b} \Phi_{\alpha\beta} \,,
\ee
whereas the antisymmetric longitudinal projection of $m_{\mu}{}^a$, given by $\tau^{\alpha[a}m_{\alpha}{}^{b]}$, can be gauged away via a $Z_{ab}$-transformation as is clear from eq.\eqref{Zabtransfoofmmua}. As such $m_{\mu}{}^a$ can be expressed in terms of $\Phi_{\alpha\beta}$.
With $\{\Gamma^i_{\alpha\beta}, \Gamma^{\e}_{\alpha\beta}\}$ being the only nonzero connection coefficients, the longitudinal components of the Ricci tensor become
\begin{align}
R_{\alpha\beta}(\Gamma) & = \Delta\Phi_{\alpha\beta} + R_{\alpha\beta}(\text{$AdS_2$}) \nonumber\\
& = \Delta \phi_{\alpha\beta} + (D-1) \Lambda \tau_{\alpha\beta}  \,,
\end{align}
where we have used that $R_{\alpha\beta}(\text{$AdS_2$}) = \Lambda \tau_{\alpha\beta}$. Therefore, the nonzero components of the Poisson equation \eqref{NCbulk} read as follows \cite{Bagchi:2009my}:
\begin{equation}\label{NCbulk}
\Delta \phi_{\alpha\beta} = \Bigl(V_{D-2}G\rho - (D-1) \Lambda\Bigr) \tau_{\alpha\beta} \,,
\end{equation}
where $D$ is the dimension of spacetime. Notice how the Laplacian on the left only contains information about the transverse space, whereas the geometry of the ${\rm AdS}_2$-foliation is only on the right hand side of \eqref{NCbulk}. This concludes our discussion of the addition of the cosmological constant to the theory.

\section{Conclusions and outlook}

We have shown how the theory of Newton-Cartan can be extended from particles moving in a flat background to strings
moving in a cosmological background. One way to obtain the desired equations corresponding to these extensions is to
gauge the transverse translations. This necessitates the introduction of a new field, which is identified as the gravitational potential. The resulting equations of motion are the ones used by a Galilean observer. Alternatively,  one can first gauge the full extended (stringy) Galilei algebra and, next, gauge-fix some of the symmetries in order to obtain the symmetries that are appropriate to a Galilean observer. The (central) extensions of the algebras involved play a crucial role in this procedure. To obtain the (stringy) Newton-Cartan theory, conventional constraints are imposed to convert the spacetime translations into general coordinate transformations and to make the spin connections dependent fields.
Further on-shell constraints are imposed  on the curvature of the transverse space and, in the string case,  on the curvature of the foliation space. The transverse space is chosen to be flat, whereas for the string the on-shell constraint on the longitudinal boost curvature can be chosen such that one obtains either a flat foliation (corresponding to the stringy Galilei group) or an ${\rm AdS}_2$-foliation (corresponding to the stringy Newton-Hooke group). The first choice describes the non-relativistic limit of a string moving in a Minkowski background, whereas the second choice describes the non-relativistic limit of a string moving in an ${\rm AdS}_D$ background. The analysis can easily be extended to arbitrary branes, in which case one should use  extended brane Galilei algebras \cite{Brugues:2006yd}.

It is interesting to compare our results with the literature on the application of Newton-Cartan theory in the non-relativistic limit of the AdS/CFT correspondence. This has been discussed in, e.g., \cite{Lin:2008pi,Brattan:2011my} where some subtleties of this application are discussed. In \cite{Bagchi:2009my} it was noted that the non-relativistic limit on the CFT-side of the correspondence should give the so-called Galilei conformal symmetry group. This Galilean conformal symmetry group is the boundary realization of the stringy Newton-Hooke algebra in the bulk \cite{Sakaguchi:2009de}. The dual gravity theory should then be a Newton-Cartan theory with an ${\rm AdS}_2$-foliation describing strings, instead of the usual $\mR$-foliation which describes particle Newton-Cartan theory. The gauging procedure outlined in this work provides the framework of developing such a theory from a gauge perspective.

It is known that the Newton-Cartan theory can be obtained from a dimensional reduction of general relativity along a null-Killing vector, see e.g.~\cite{Duval:1984cj,Julia:1994bs}.\,\footnote{In
\cite{Julia:1994bs} also a proposal for an action describing the NC bulk dynamics has been made.
For AdS/CFT applications this is a desirable feature.} The central charge gauge field $m_\mu$ is related to the Kaluza-Klein vector corresponding to this null direction.
It would be interesting to investigate if the stringy version of the Newton-Cartan theory presented in this paper can also be obtained by a  null-reduction from higher dimensions such that the deformation potentials $m_\mu{}^a$ and $m_\mu{}^{ab}$ obtain a similar Kaluza-Klein interpretation.
This possibility should be related to the fact that the extended Newton-Hooke p-brane algebra in D dimensions is a subalgebra of the "multitemporal" conformal group SO(D+1,p+2) in one dimension higher  \cite{Brugues:2006yd}.

One way to obtain null-directions  is to start from a relativistic string coupled to a constant B-field with vanishing  field strength and to T-dualize this string  along its spatial world-sheet direction and perform the non-relativistic limit.
The T-dual picture is a pp-wave which has a null-direction \cite{Gomis:2005pg}. One could now use this null direction for a Kaluza-Klein reduction along the lines of \cite{Julia:1994bs} and see whether one obtains the stringy Newton-Cartan theory constructed in this paper.

Finally, an interesting extension of the stringy Newton-Cartan theory would be to apply the gauging procedure as presented here to the supersymmetric extension of the stringy Galilei algebra \cite{Gomis:2005pg}. We hope to return to these issues in the nearby future.

\section*{Acknowledgements}

We wish to thank J. Rosseel for useful discussions. E.B. wishes to thank the Universitat de Barcelona and the Isaac Newton institute, where part of this work was done, for its hospitality. E.B. and J.G want to thank the organizers of the CECS workshop of 2010 in Valdivia, where part of this work originated. The work of R. Andringa is supported by an Ubbo Emmius Fellowship. We also acknowledge partial financial support from projects FP2010-20807-C02-01, 2009SGR502 and CPAN Consolider CSD 2007-00042.

\begin{appendix}

\section{Notation and conventions}
\label{appendixnotation}
Our notation and conventions are as follows. For the metric the mostly-plus convention is taken. A positive cosmological constant $\Lambda > 0$ describes a deSitter space, whereas $\Lambda < 0$ describes an anti deSitter space.

Flat target-space indices are given by $A = \{a,a'\}$, where $\{a\}$ is longitudinal and $\{a'\}$ is transverse, e.g.
\be
\zeta^A = \{ \zeta^{a}, \zeta^{a'} \} \,.
\ee
For a particle we write $\{a=\underline{0}\}$ and $\{a'=1,\ldots,D-1\}$, whereas for a string we write $\{a=\underline{0},\underline{1}\}$ and $\{a'=2\ldots D-1\}$. Curved target-space indices are given by $\mu=\{\alpha,i\}$, where $\{\alpha\}$ is longitudinal and $\{i\}$ is transverse, e.g.
\be
\xi^{\mu} = \{ \xi^{\alpha}, \xi^{i} \} \,.
\ee
For a particle we write $\{\alpha=0\}$ and $\{i=1,\ldots,D-1\}$, and for a string we write $\{\alpha=0,1\}$ and $\{i=2, \ldots ,D-1\}.$ Finally, we indicate world-sheet indices with $\{\bar{\alpha},\bar{\beta},\ldots\}$, and the world-sheet coordinates as $\{\sigma^{\bar{\alpha}}\}$.

\section{The extended stringy Galilei algebra}
\label{ExtStringGalalg}
We associate the following generators to the symmetries of the extended stringy Galilei algebra \cite{Brugues:2004an}
:
\begin{eqnarray}\label{generators}
H_a&:&\ \ \ \text{longitudinal translations}\nonumber \\[.1truecm]
P_{a'}&:&\ \ \ \text{transverse  translations}\nonumber \\[.1truecm]
M_{ab} &:&\ \ \ \text{longitudinal Lorentz transformations} \\[.1truecm]
M_{a'a}&:&\ \ \ \text{``boost'' transformation}\nonumber \\[.1truecm]
M_{a'b'}&:&\ \ \ \text{transverse rotations}\nonumber \\[.1truecm]
Z_a\,, Z_{ab}&:&\ \ \ \text{extended  transformations}\,,\nonumber
\end{eqnarray}
with $Z_{ab}=-Z_{ba}$.

The nonzero commutators of the un-deformed stringy Galilei  algebra read
\begin{align}
[M_{b'c} , H_a]  & = \eta_{ac}P_{b'}\,, & [M_{b'c'}, P_{a'}]  & = -2\eta_{a'[b'}P_{c']} \,, \nonumber\label{AdS2subalgebra} \\[.1truecm]
[M_{c'd}, M_{ef}] & = 2 \eta_{d[e}M_{|c'|f]}\,, & [M_{c'd'}, M_{e'f}] & = -2 \eta_{e'[c'}M_{d']f} \,,\\[.1truecm]
[M_{c'd'},M_{e'f'}]  & = 4 \eta_{[c' [e'}M_{f']d']} &[M_{bc} , H_a]  & = -2\eta_{a[b}H_{c]}\,,\nonumber
\end{align}
where $a=\underline{0},\underline{1}$ are the two longitudinal foliating directions and $a^\prime = 2,\cdots ,D-1$ are the $D-2$ transverse directions. Note that the Lorentz algebra $\mathfrak{so}(1,1)$ of the two-dimensional foliation space is Abelian while for general p-branes, where the symmetries of the foliation space are generated by the algebra $\mathfrak{so}(1,p)$, this would not be the case. The extensions suggested by the Poisson brackets corresponding to the non-relativistic string action \eqref{nonrelaction} are given by \cite{Brugues:2006yd}
\begin{align}
[P_{a'}, M_{b'b}] & = \eta_{a'b'}Z_b\,, & [M_{a'a}, M_{b'b}] & = -\eta_{a'b'}Z_{ab} \,, \nonumber\\[.1truecm]
[H_a, Z_{bc}] & = 2\eta_{a[b}Z_{c]}\,,  & [Z_{ab}, M_{cd}] & = 4 \eta_{[a[c}Z_{d]b]}\,, \label{NHdeformation} \\[.1truecm]
[Z_a, M_{bc}] & = 2\eta_{a[b}Z_{c]}\,. \nonumber
\end{align}

The gauge transformations of the gauge fields \eqref{gaugefields} corresponding to the generators \eqref{generators} of the deformed stringy Galilei  algebra are given by
\begin{eqnarray}
\delta \tau_{\mu}{}^a & =& \partial_{\mu} \tau^a - \tau^b \omega_{\mu}{}^{ab} + \lambda^{ab}\tau_{\mu}{}^b\,, \nonumber\\ [.15truecm]
\delta e_{\mu}{}^{a'} & = &\partial_{\mu} \zeta^{a'} - \zeta^{b'}\omega_{\mu}{}^{a'b'}  + \lambda^{a'b'}e_{\mu}{}^{b'}
+ \lambda^{a'a}\tau_{\mu}{}^a - \tau^a \omega_{\mu}{}^{a'a}\,, \nonumber\\ [.15truecm]
\delta \omega_{\mu}{}^{ab} & =& \partial_{\mu} \lambda^{ab}\,, \nonumber\\ [.15truecm]
\delta \omega_{\mu}{}^{a'a} & =& \partial_{\mu} \lambda^{a'a} - \lambda^{a'b}\omega_{\mu}{}^{ab} + \lambda^{ab}\omega_{\mu}{}^{a'b} + \lambda^{a'b'}\omega_{\mu}{}^{b'a} - \lambda^{b'a}\omega_{\mu}{}^{a'b'}\,, \\ [.15truecm]
\delta \omega_{\mu}{}^{a'b'} & =& \partial_{\mu} \lambda^{a'b'} + 2 \lambda^{c'[a'}\omega_{\mu}{}^{b']c'}\,, \nonumber\\ [.15truecm]
\delta m_{\mu}{}^a & =& \partial_{\mu} \sigma^a + \lambda^{a'a}e_{\mu}{}^{a'} - \zeta^{a'}\omega_{\mu}{}^{a'a} + \lambda^{ab}m_{\mu}{}^b - \sigma^b \omega_{\mu}{}^{ab} + \tau^b m_{\mu}{}^{ab} - \sigma^{ab} \tau_{\mu}{}^b \,,\nonumber\\ [.15truecm]
\delta m_{\mu}{}^{ab} & =& \partial_{\mu} \sigma^{ab}
- \lambda^{a'a}\omega_{\mu}{}^{a'b} + \lambda^{a'b}\omega_{\mu}{}^{a'a}+ \sigma^{c[a}\omega_{\mu}{}^{b]c} + \lambda^{c[a}m_{\mu}{}^{b]c} \,,\nonumber
\label{NHgaugetransformations}
\end{eqnarray}
where we have used the gauge parameters \eqref{parameters}.
The corresponding gauge-invariant curvatures are given by\footnote{For general p-branes we would have $\delta \omega_{\mu}{}^{ab}  = \partial_{\mu} \lambda^{ab} + 2 \lambda^{c[a}\omega_{\mu}{}^{b]c}$ and \\ $R_{\mu\nu}{}^{ab}(M) = 2 \Bigl( \partial_{[\mu}\omega_{\nu]}{}^{ab} - \omega_{[\mu}{}^{ca}\omega_{\nu]}{}^{bc}\Bigr)$.}
\begin{eqnarray}\label{curvatures}
R_{\mu\nu}{}^{a}(H) & =& 2 D_{[\mu}\tau_{\nu]}{}^a \,, \nonumber\\[.15truecm]
R_{\mu\nu}{}^{a'}(P) & =& 2 \Bigl(D_{[\mu}e_{\nu]}{}^{a'} - \omega_{[\mu}{}^{a'a}\tau_{\nu]}{}^{a} \Bigr)\,, \nonumber\\[.15truecm]
R_{\mu\nu}{}^{ab}(M) & = &2\,  \partial_{[\mu} \omega_{\nu]}{}^{ab}  \,, \nonumber\\[.15truecm]
R_{\mu\nu}{}^{a'a}(M') & = &2\,  D_{[\mu} \omega_{\nu]}{}^{a'a} \,, \\[.15truecm]
R_{\mu\nu}{}^{a'b'}(M'') & = &2 \Bigl( \partial_{[\mu}\omega_{\nu]}{}^{a'b'} - \omega_{[\mu}{}^{c'a'}\omega_{\nu]}{}^{b'c'}\Bigr)\,, \nonumber\\[.15truecm]
R_{\mu\nu}{}^{a}(Z) & = &2 \Bigl( D_{[\mu} m_{\nu]}{}^a  + e_{[\mu}{}^{a'}\omega_{\nu]}{}^{a'a} - \tau_{[\mu}{}^b m_{\nu]}{}^{ab} \Bigr) \,, \nonumber\\[.15truecm]
R_{\mu\nu}{}^{ab}(Z) & =& 2 \Bigl( D_{[\mu} m_{\nu]}{}^{ab}
+ \omega_{[\mu}{}^{a'a}\omega_{\nu]}{}^{a'b} \Bigr) \,, \nonumber\label{NHcurvatures}
\end{eqnarray}
where $M$, $M'$ and $M''$ indicate the generators corresponding to  longitudinal Lorentz transformations,
``boost'' transformations and transverse rotations, respectively. The derivative $D_{\mu}$ is covariant with respect to these  $M$, $M'$ and $M''$ transformations.

Finally, the curvatures \eqref{curvatures} satisfy the Bianchi identities
\begin{align}\label{Bianchi}
D_{[\rho}R_{\mu\nu]}{}^{a} (H) & = - R_{[\rho\mu}{}^{ab}(M)\tau_{\nu]}{}^b  \,, \nonumber\\[.15truecm]
D_{[\rho}R_{\mu\nu]}{}^{a'} (P) & = -R_{[\rho\mu}{}^{a'b'}(M'')e_{\nu]}{}^{b'} - R_{[\rho\mu}{}^{a'a}(M')\tau_{\nu]}{}^a    \,,      \nonumber\\[.15truecm]
D_{[\rho}R_{\mu\nu]}{}^{ab} (M) & = 0 \,,        \nonumber\\[.15truecm]
D_{[\rho}R_{\mu\nu]}{}^{a'a} (M') & = - R_{[\rho\mu}{}^{ab}(M)\omega_{\nu]}{}^{a'b} - R_{[\rho\mu}{}^{a'b'}(M'')\omega_{\nu]}{}^{b'a} \,, \\[.15truecm]
D_{[\rho}R_{\mu\nu]}{}^{a'b'} (M'') & = 0  \,,\nonumber \\[.15truecm]
D_{[\rho} R_{\mu\nu]}{}^{a}(Z) & = - R_{[\rho\mu}{}^{ab}(M)m_{\nu]}{}^b + R_{[\rho\mu}{}^{a'}(P)\omega_{\nu]}{}^{a'a}
- R_{[\rho\mu}{}^{a'a}(M')e_{\nu]}{}^{a'} \,, \nonumber\\[.15truecm]
& - R_{[\rho\mu}{}^{a}(H)m_{\nu]}{}^{ab} + R_{[\rho\mu}{}^{ab}(Z)\tau_{\nu]}{}^b \,, \nonumber \\[.15truecm]
D_{[\rho} R_{\mu\nu]}{}^{ab}(Z) & = R_{[\rho\mu}{}^{c[a}(M)m_{\nu]}{}^{b]c} + R_{[\rho\mu}{}^{a'a}(M')\omega_{\nu]}{}^{a'b} - R_{[\rho\mu}{}^{a'b}(M')\omega_{\nu]}{}^{a'a} \nonumber \,.
\end{align}

\section{Newton-Cartan geodesic equations}
\label{NCgeodesicdetails}
Here we give some details about the derivation of the geodesic equations \eqref{geodesiceq} and \eqref{NCgeodesicstring}. We start with the point particle case. For that purpose we write the Lagrangian \eqref{altparticleaction} as
\begin{align}
L & = \frac{m}{2}N^{-1}\dot{x}^{\mu}\dot{x}^{\nu}\Bigl(h_{\mu\nu} - 2m_{\mu}\tau_{\nu}\Bigr) \nonumber\\
 & \equiv \frac{m}{2}N^{-1}\dot{x}^{\mu}\dot{x}^{\nu}H_{\mu\nu} \,, \label{particleactionappendix}
\end{align}
where we defined
\be
H_{\mu\nu} \equiv h_{\mu\nu} - 2m_{(\mu}\tau_{\nu)}, \ \ \ \ N \equiv \tau_{\mu}\dot{x}^{\mu} \,.
\ee
Varying the Lagrangian \eqref{particleactionappendix} with respect to $\{x^{\lambda}\}$ and using the metric compatibility condition $\p_{[\mu}\tau_{\nu]}=0$ gives
\begin{align}
-Nm^{-1}\frac{\delta L}{\delta x^{\lambda}}& = \Bigl( N^{-2}\dot{N}\tau_{\lambda} H_{\mu\nu} - \frac{1}{2}N^{-1} \tau_{\lambda} \p_{\rho} H_{\mu\nu}\dot{x}^{\rho} - \frac{1}{2} \p_{\lambda} H_{\mu\nu} + \p_{\nu} H_{\mu\lambda}  \Bigr) \dot{x}^{\mu} \dot{x}^{\nu} \nonumber\\
&- N^{-1} \tau_{\lambda} H_{\mu\nu}\dot{x}^{\mu} \ddot{x}^{\nu}
 - N^{-1} \dot{N} H_{\mu\lambda}\dot{x}^{\mu} + H_{\mu\lambda}\ddot{x}^{\mu} = 0 \,. \label{Lfunctionalderivative}
\end{align}
First, we contract this equation with $h^{\lambda\sigma}$. This gives
\be
h^{\lambda\sigma}\Bigl(\p_{\nu}H_{\mu\lambda} - \frac{1}{2}\p_{\lambda}H_{\mu\nu} \Bigr)\dot{x}^{\mu} \dot{x}^{\nu} + h^{\lambda\sigma}H_{\mu\lambda}\ddot{x}^{\mu} - N^{-1}\dot{N}h^{\lambda\sigma}H_{\mu\lambda}\dot{x}^{\mu} = 0 \,. \label{Geohcontraction}
\ee
One can now use the Newton-Cartan metric relations \eqref{NCvielbeinrelations}, $\p_{[\mu}\tau_{\nu]}=0$ and
\be
\dot{N} = \tau_{\mu}\ddot{x}^{\mu} + \p_{\mu}\tau_{\nu} \dot{x}^{\mu} \dot{x}^{\nu} \,.
\ee
Some manipulation then shows that \eqref{Geohcontraction} gives the geodesic equation \eqref{geodesiceq},
\be
{\ddot x}^\mu +\Gamma^\mu_{\nu\rho}{\dot x}^\nu {\dot x}^\rho = \frac{\dot{N}}{N}\dot{x}^{\mu} \,, \label{geodesiceqAppendix}
\ee
with the connection given by \eqref{NCparticleConhtauK}. Second, one can contract \eqref{Lfunctionalderivative} with $\tau^{\lambda}$. The resulting expression contains, among others, terms proportional to $\ddot{x}^{\mu}$. If one uses the geodesic equation \eqref{geodesiceqAppendix} to rewrite these in terms of $\dot{x}^{\mu}$ one can finally show that this $\tau^{\lambda}$-contraction of \eqref{Lfunctionalderivative} is trivially satisfied. \\

The calculation concerning the string Lagrangian \eqref{altstringaction} leading to the stringy geodesic equation \eqref{NCgeodesicstring} can be done in a similar way.
We first write
\be
H_{\mu\nu} = h_{\mu\nu} - 2m_{(\mu}{}^a\tau_{\nu)}{}^a \,,
\ee
such that \eqref{altstringaction} becomes
\begin{equation}
L = - \frac{T}{2} \sqrt{-\text{det}(\tau)} \tau^{\bar{\alpha}\bar{\beta}}\p_{\bar{\alpha}} x^{\mu} \p_{\bar{\beta}} x^{\nu} H_{\mu\nu} \,. \label{altaltstring}
\end{equation}
We next use the relations
\begin{align}
\delta \sqrt{-\text{det}(\tau)} & = \frac{1}{2}\sqrt{-\text{det}(\tau)} \tau^{\bar{\alpha}\bar{\beta}}\delta \tau_{\bar{\alpha}\bar{\beta}} \,, \nonumber\\
\delta\tau^{\bar{\alpha}\bar{\beta}} & = -\tau^{\bar{\alpha}\bar{\gamma}}\tau^{\bar{\beta}\bar{\e}} \delta \tau_{\bar{\gamma}\bar{\e}} \,, \nonumber\\
\delta \tau_{\bar{\alpha}\bar{\beta}} & = 2 \p_{\bar{\alpha}} x^{\mu} \p_{\bar{\beta}} \delta x^{\lambda} \tau_{\mu\lambda} + \p_{\bar{\alpha}} x^{\mu} \p_{\bar{\beta}} x^{\nu} \p_{\lambda}\tau_{\mu\nu} \delta x^{\lambda} \,, \nonumber\\
\p_{\bar{\alpha}} \Bigl(\sqrt{-\text{det}(\tau)} \tau^{\bar{\alpha}\bar{\beta}} \p_{\bar{\beta}} x^{\mu} \Bigr)  & = \sqrt{-\text{det}(\tau)} \tau^{\bar{\alpha}\bar{\beta}}\na_{\bar{\alpha}} \p_{\bar{\beta}} x^{\mu}\,,   \nonumber\\
\p_{\rho}\tau_{\mu\nu}  + \p_{\mu}\tau_{\rho\nu} - \p_{\nu}\tau_{\rho\mu} & = \Gamma^{\lambda}_{\mu\rho} \tau_{\lambda\nu} \,,
\end{align}
where the last identity follows from the metric compatibility condition $\na_{\rho}\tau_{\mu\nu}=0$. Varying \eqref{altaltstring} with respect to $\{x^{\lambda}\}$ now gives the geodesic equation \eqref{NCgeodesicstring},
\be
\tau^{\bar{\alpha}\bar{\beta}} \Bigl(\na_{\bar{\alpha}}\,\p_{\bar{\beta}}\,x^{\rho} + \p_{\bar{\alpha}}x^{\mu} \p_{\bar{\beta}}x^{\nu}\, \Gamma^{\rho}_{\mu\nu} \Bigr) = 0 \,,
\ee
with the connection $\Gamma^{\rho}_{\mu\nu}$ given by \eqref{NHcovariantconnection}. This connection is equivalent to the connection \eqref{NHGammastr} given by the vielbein postulates.

\section{Some properties of ${\rm AdS}_2$}
\label{ads2appendix}
In terms of coordinates $x^\alpha = \{t,z\}$ we write the ${\rm AdS}_2$-metric as $\tau_{\alpha\beta}$, and the corresponding line interval as
\begin{equation}
ds^2 =  - \Bigl(1 + \frac{z^2}{R^2}	\Bigr)dt^2 + \Bigl(1 + \frac{z^2}{R^2}	 \Bigr)^{-1}dz^2 \,,
\end{equation}
where $R$ is the radius of curvature. The  nonzero Christoffel components in this coordinate system
are given by
\begin{equation}
\Gamma^z_{tt} = z\Bigl(\frac{z^2 + R^2}{R^4}\Bigr), \ \ \ \Gamma^z_{zz} = \frac{-z}{z^2 + R^2}, \ \ \ \Gamma^t_{zt} = \frac{z}{z^2 + R^2} \,. \label{AdS2Christoffels}
\end{equation}

The three isometries of the ${\rm AdS}_2$-space parametrized by $\{\zeta^0, \zeta^1, \lambda^{01} \}$ are described by the Killing vectors\footnote{Notice that $k_{\{02\}}$ describes the fact that the $AdS_2$ metric is static. We could rescale the time coordinate $t$ with $R$ to get $k_{\{02\}} = - \p_t$.}
\begin{align}
k_{\{01\}} & = \frac{zR\cos{\frac{t}{R}}}{\sqrt{z^2 + R^2}} \p_t + \sqrt{z^2 + R^2}\sin{\frac{t}{R}}\p_z \,, \nonumber\\
k_{\{02\}} & = -R \p_t \,, \nonumber\\
k_{\{12\}} & = \frac{Rz \sin{\frac{t}{R}}}{\sqrt{z^2 + R^2}}\p_t - \sqrt{z^2 + R^2} \cos{\frac{t}{R}}\p_z \,. \label{AdS2Killings}
\end{align}
One can check that these vectors indeed form an $\mathfrak{so}(2,1)$ algebra, and that the components of the vectors \eqref{AdS2Killings} obey the Killing equation
\be
\LL_k \tau_{\alpha\beta} = 2\na_{(\alpha}k_{\beta)} = 0 \,.
\ee
Acting with the Killing vectors \eqref{AdS2Killings} on the coordinates $x^\alpha = \{t,z\}$ induces the infinitesimal isometry transformations
\begin{eqnarray}
\delta_H t & =& \zeta^0 - \zeta^1 \frac{z}{\sqrt{z^2 + R^2}}\sin{(\frac{t}{R})}\,,\nonumber\\ [.2truecm]
\delta_H z &=& \zeta^1 \frac{\sqrt{z^2 + R^2}}{R} \cos{(\frac{t}{R})}\,, \\ [.2truecm]
\delta_M t & = &\lambda^{01} \frac{ z R\cos{(\frac{t}{R})}}{\sqrt{z^2 + R^2}}\nonumber\,,\\ [.2truecm]
\delta_M z &=& -\lambda^{01} \sqrt{z^2 + R^2} \sin{(\frac{t}{R})} \,. \nonumber  \label{inftz}
\end{eqnarray}
Note that in the limit $R\rightarrow\infty$ these rules reduce to the stringy Galilei ones given by
$\xi^\alpha(x^\alpha) = \lambda^\alpha{}_\beta x^\beta + \zeta^\alpha$ which are the isometries of a flat
$M_{1,1}$ foliation space.

\end{appendix}

\end{document}